\documentclass[10pt,a4paper,final]{iopart}

\usepackage{iopams}

\usepackage{hyperref}
\usepackage{geometry}
\usepackage{graphicx}
\usepackage{caption}
\usepackage{subcaption}
\usepackage{qcircuit}
\usepackage{bm}
\usepackage{braket}
\usepackage{amsthm}
\newcommand{\argmax}{\mathop{\mathrm{argmax}}}
\usepackage{xcolor}

\begin{document}

\title{Quantum information criteria for model selection in quantum state estimation}

\author{Hiroshi Yano$^1$ \& Naoki Yamamoto$^{1,2}$}

\address{$^1$ Department of Applied Physics and Physico-Informatics, Keio University,
Hiyoshi 3-14-1, Kohoku, Yokohama 223-8522, Japan}
\address{$^2$ Quantum Computing Center, Keio University, Hiyoshi 3-14-1, Kohoku,
Yokohama 223-8522, Japan}
\ead{hiroshi.yano.gongon@keio.jp}
\vspace{10pt}
% \begin{indented}
% \item[]August 2017
% \end{indented}

\begin{abstract}
  Quantum state estimation (or state tomography) is an indispensable task in quantum 
  information processing. 
  Because full state tomography that determines all elements of the density matrix is 
  computationally demanding, one usually takes the strategy of assuming a certain model 
  of quantum states and identifying the model parameters. 
  However, it is difficult to make a valid assumption given little prior knowledge on 
  a quantum state of interest, and thus we need a reasonable \textit{model selection} 
  method for quantum state estimation. 
  Actually, in the classical statistical estimation theory, several types of 
  information criteria have been established and widely used in practice for 
  appropriately choosing a classical statistical model. 
  In this study, we propose quantum information criteria for evaluating the 
  quality of the estimated quantum state in terms of the quantum relative entropy, 
  which is a natural quantum analogue of the classical information criterion 
  defined in terms of Kullback–Leibler divergence. 
  In particular, we derive two quantum information criteria depending on the type 
  of an estimator for the quantum relative entropy; one uses the log-likelihood and 
  the other uses the classical shadow. 
  The general role of information criteria is to predict the performance of an estimated 
  model for unseen data, although it is a function of only sampled data; this 
  generalization capability of the proposed quantum information criteria is 
  evaluated in numerical simulations. 
\end{abstract}

%\maketitle

%%%%%%%%%%%%%%%%%%%%%%%%%%%%%%%%%%%%%%%%%%%%%%%%%%
\section{Introduction}\label{sec:intro}
%%%%%%%%%%%%%%%%%%%%%%%%%%%%%%

Quantum state estimation (or state tomography) \cite{paris2004Quantum} is an indispensable 
task in quantum information processing.
Its purpose is to reconstruct a density matrix from the measurement data on an unknown 
quantum state obtained through an experiment. 
Generally, the increase of the size of quantum systems requires an exponential amount 
of measurement data in the system size for the full state tomography. 
Therefore, the state estimation for large quantum systems is possible only when some 
prior knowledge on the state is available and an associated specific technique requiring 
a feasible number of measurements can be constructed. 
The compressed sensing, for instance, is an efficient approach when a density matrix of 
the target state is of low rank \cite{gross2010Quantum}. 
Also, the matrix product state is a promising tool for estimating low-entangled states, 
such as ground states of certain local Hamiltonians in condensed matter physics 
\cite{cramer2010Efficient,lanyon2017Efficient}. 
However, these approaches are not state universal (that is, the algorithms cannot be 
applied to arbitrary quantum states), and we need a model that can handle a wider class 
of quantum states. 
Promising such models are neural-network quantum states \cite{carleo2017Solving} and 
quantum Boltzmann machines (QBMs) \cite{amin2018Quantum}, which both offer a compact 
variational representation of many-body quantum states. 
Note that those models are further characterized by some hyperparameters such as 
the network width/depth of neural-network quantum states and the number of hidden 
variables of QBMs. 
Therefore, with a limited number of measurement data, we need to determine the best 
model of a quantum state that can explain the measurement data and moreover well predict 
some features associated with unseen data.

In the classical statistics, the theory for model selection \cite{burnham2004Model} 
has been extensively developed, ranging from hypothesis testing to cross validation. 
In 1973, Akaike developed a novel paradigm for model selection \cite{akaike1973information}. 
That is, based on the likelihood principle and Kullback-Leibler (KL) divergence, he 
derived the quantity for model selection called Akaike information criterion (AIC), 
which enables us to identify the best model of a target probability distribution using 
given empirical data. 
Since then, several other information criteria have been derived including Takeuchi 
information criterion (TIC) \cite{takeuchi1976Distribution}, to deal with models 
in various scenarios in social science as well as natural science 
\cite{bozdogan2000Akaike,parzen1998Selected}.

In the quantum state estimation, on the other hand, only a few works have been done 
for the problem of model selection. 
Some researches used AIC to select the best one from multiple candidates of 
parameterized quantum states \cite{usami2003Accuracy,yin2011Information,schwarz2013Error}.
In Ref.~\cite{usami2003Accuracy}, AIC was introduced to eliminate redundant parameters 
of the parametric model of quantum states and succeeded in enhancing the estimation 
accuracy. 
It was shown there that reducing the number of parameters by AIC also contributed 
to circumventing the problem of local minima of the likelihood function in finding 
the maximum of the likelihood function. 
In Ref.~\cite{yin2011Information}, AIC was used to determine the structure of the 
target state for realizing quantum state estimation on a large scale.
The model selection by AIC can be further applied to identify a noise model 
\cite{schwarz2013Error}. 
Note that these works are fully formulated within the classical statistics, where 
the KL divergence for classical probability distributions and accordingly AIC are 
calculated. 
Because these quantities are defined with fixed measurement process, such a classical 
approach does not fully extract information from the target quantum system for the 
best model selection.

Therefore, for the purpose of developing a genuine model selection method for 
quantum systems, in this paper we propose a new quantum-oriented information 
criterion that evaluates a quantum state in terms of the quantum relative entropy 
instead of the KL divergence. 
Actually, because the quantum relative entropy itself is independent of measurements, 
the proposed criterion can measure the intrinsic validity of the model. 
Note that we actually find some recent works that use the quantum relative entropy 
in quantum estimation problems 
\cite{kieferova2017Tomography,wiebe2019Generative,kappen2020Learning}; 
for instance, Ref.~\cite{kieferova2017Tomography} used the quantum relative entropy 
to train QBM for learning quantum states.

Here we briefly explain the essence of the proposed information criterion. 
In the classical case, the information criteria are constructed by using i.i.d. samples 
$\bm{x}_n = \{x_1, x_2, ..., x_n\}$ generated from an unknown true system. 
In particular, AIC uses this dataset to calculate the maximum likelihood estimator 
and evaluates the KL divergence between the estimated statistical model and the target 
true distribution. 
In order to reflect the concept of i.i.d. sampling for the quantum state estimation, 
we assume that a finite number of data are obtained through a tomographically complete 
measurement with uniform weights on a true quantum state; 
that is, we apply a unitary operator randomly chosen from the ensemble of unitaries 
and then measure the rotated state in the computational basis, the $n$ repetitions 
of which give us the dataset $\bm{x}_n = \{x_1, x_2, ..., x_n\}$. 
We also introduce an alternative representation of the measurement outcomes 
$\hat{\bm\rho}_n = \{\hat{\rho}_1, \hat{\rho}_2, ..., \hat{\rho}_n\}$ using 
the classical shadow \cite{huang2020Predicting}. 
Note that $\hat{\rho}_\alpha$ corresponds to $x_\alpha$ for $\alpha = 1,...,n$, but 
we use these two types of measurement outcomes to define two different quantum 
information criteria.

The paper is organized as follows. 
In Section~\ref{sec:InformationCriteria}, we provide a brief overview of classical 
information criteria. 
Section~\ref{sec:quantumInformationCriteria} derives the proposed quantum information 
criteria, which is a quantum generalization of the classical information criteria. 
In particular, we considered two different approaches to compute the quantum information 
criteria: one focusing on the log-likelihood in Section~\ref{subsec:qic_ll} and the other 
focusing on the classical shadow in Section~\ref{subsec:qic_shadow}. 
In Section~\ref{sec:numDemo}, we present a numerical demonstration of calculating the 
quantum information criteria for selecting a better quantum model and provide arguments 
on the practical limitation of their computation. 
Finally, Section~\ref{sec:conclusion} concludes this paper.

%%%%%%%%%%%%%%%%%%%%%%%%%%%%%%%%%%%%%%%%%%%%%%%%%%
\section{Classical information criteria: AIC and TIC}
\label{sec:InformationCriteria}

In this section, we briefly describe two information criteria for classical statistical 
models, AIC \cite{akaike1973information,akaike1974New} and TIC 
\cite{takeuchi1976Distribution}, both of which are the measure based on the KL divergence. 
For readers who are interested in the detailed derivation of AIC and TIC, we refer to 
Ref.~\cite{burnham2004Model,konishi2008information}.

Assume that the data $\bm{x}_n = \{x_1, x_2, ..., x_n\}, \, x_\alpha \in \mathbb{R}$ are 
generated from an unknown true model with probability density function $g(x)$. 
We estimate this target distribution by using the maximum likelihood method with 
a parametric model $\{ f(x|\bm\theta) : \bm\theta \in \Theta \subset \mathbb{R}^p \}$ 
having a $p$-dimensional vector of parameters. 
The estimated model with the maximum likelihood estimator $\hat{\bm\theta}$ is evaluated by measuring 
the KL divergence between the true distribution and the model $f(x|\hat{\bm\theta})$; 
\begin{equation}\label{eq:KL}
  KL(g(z) \| f(z|\hat{\bm\theta})) = \mathbb{E}_{g(z)} \left[ \log g(Z) \right] - \mathbb{E}_{g(z)} \left[ \log f(Z|\hat{\bm\theta}) \right].
\end{equation}
The KL divergence is $0$ when $f=g$, and positive otherwise.
Because the first term of the KL divergence is independent of the model $f$, for 
the purpose of comparing some candidate models for a certain true distribution, it is 
enough to evaluate the second term, i.e., the expected log-likelihood or the negative 
classical cross entropy 
\begin{equation}\label{eq:negativeCCE}
  H(g(z)\|f(z|\hat{\bm\theta})) = \mathbb{E}_{g(z)} \left[ \log f(Z|\hat{\bm\theta}) \right]. 
\end{equation}
A good estimator for the expected log-likelihood turns out to be 
\begin{equation}
\label{eq:est_expected_loglikelihood}
  \mathbb{E}_{\hat{g}(z)} \left[ \log f(Z|\hat{\bm\theta}) \right] = \frac{1}{n} \sum_{\alpha=1}^n \log f(x_\alpha|\hat{\bm\theta}) = \frac{1}{n} \ell(\hat{\bm\theta}),
\end{equation}
where $\ell(\hat{\bm\theta}) = \sum_{\alpha=1}^n \log f(x_\alpha|\hat{\bm\theta})$ 
is the log-likelihood and $\hat{g}(z) = (1/n) \sum_{\alpha=1}^n \delta(z - x_\alpha)$ 
is an empirical probability density function ($\delta(z)$ is the delta function). 
Note that the above estimator is a biased estimator; hence we correct the bias, and 
the resulting modified estimator is the information criterion. 
In the general setting under the regularity assumption, the following information 
criterion TIC was derived:
\begin{equation}
\label{eq:TIC}
  \mathrm{TIC} = -2 \ell(\hat{\bm\theta}) + 2 \Tr\left( I(\hat{\bm\theta}) J(\hat{\bm\theta})^{-1} \right)
\end{equation}
with the conventional factor $2$, where $I(\hat{\bm\theta})$ and $J(\hat{\bm\theta})$ 
are the $p \times p$ matrices given by
\begin{eqnarray}
  I_{ij}(\hat{\bm\theta}) &= \frac{1}{n} \sum_{\alpha=1}^{n} \left. \frac{\partial \log f(x_\alpha|\bm\theta)}{\partial \theta_i} \frac{\partial \log f(x_\alpha|\bm\theta)}{\partial \theta_j} \right|_{\bm\theta = \hat{\bm\theta}}, \\
  J_{ij}(\hat{\bm\theta}) &= - \frac{1}{n} \sum_{\alpha=1}^{n} \left. \frac{\partial^2 \log f(x_\alpha|\bm\theta)}{\partial \theta_i \partial \theta_j} \right|_{\bm\theta = \hat{\bm\theta}}. 
\end{eqnarray}
Furthermore, if the parametric model includes the true distribution, i.e., 
$g(x) \in \{ f(x|\bm\theta) : \bm\theta \in \Theta\}$, the relation 
$I(\hat{\bm\theta})=J(\hat{\bm\theta})$ holds and $\Tr\left( I(\hat{\bm\theta}) J(\hat{\bm\theta})^{-1} \right)$ is equal to the number of parameters, $p$. 
This leads to the information criterion AIC:
\begin{equation}\label{eq:AIC}
  \mathrm{AIC} = -2 \ell(\hat{\bm\theta}) + 2p.
\end{equation}
Hence, AIC enables us to select a valid model, by evaluating the trade-off between 
the fitting-level of the model to the data (the first term) and the model complexity 
(the second term). 
More precisely, assume that we have two parametric models 
$\{ f_1(x|\bm\theta_1) : \bm\theta_1 \in \Theta_1 \subset \mathbb{R}^{p_1} \}$ and 
$\{ f_2(x|\bm\theta_2) : \bm\theta_2 \in \Theta_2 \subset \mathbb{R}^{p_2} \}$, with 
$p_1 < p_2$. 
After estimating the parameters and computing AIC for each model, we can validly 
select the one that takes a smaller value of AIC.

Here we remark that AIC and TIC multiplied by $-1/(2n)$ are asymptotically unbiased 
estimators of the expected log-likelihood as follows: 
\begin{eqnarray*}
  \mathbb{E}_{g(\bm{x}_n)}\left[ - \frac{1}{2n} \mathrm{TIC} \right] = \mathbb{E}_{g(\bm{x}_n)}\left[ \mathbb{E}_{g(z)} \left[ \log f(Z|\hat{\bm\theta}) \right] \right] + o\left( n^{-1} \right), \\
  \mathbb{E}_{g(\bm{x}_n)}\left[ - \frac{1}{2n} \mathrm{AIC} \right] = \mathbb{E}_{g(\bm{x}_n)}\left[ \mathbb{E}_{g(z)} \left[ \log f(Z|\hat{\bm\theta}) \right] \right] + o\left( n^{-1} \right),
\end{eqnarray*}
where the expectation $\mathbb{E}_{g(\bm{x}_n)}$ is taken with respect to the joint 
probability density function, $g(\bm{x}_n)=\prod_{\alpha=1}^{n} g(x_\alpha)$. 
Also $o\left( \cdot \right)$ means that
$\lim_{n \to \infty} n \cdot o\left( n^{-1} \right) = 0$. 
That is, both AIC and TIC stochastically fluctuate due to the finite number of data.
However, it is known that, in the case of a hierarchical series of models 
$M_1 \subset M_2 \subset \cdots$, it has no effect on comparing the model's performance 
because the fluctuations are common between models and hence they cancel out with each 
other \cite{murata1994Network}.

Here we give some remarks. 
First, we note that AIC and TIC cannot be used for singular models.
In regular models, the Fisher information matrix does exist and it is positive definite.
Thus, the Cram\'er-Rao framework holds in the asymptotic regime; as a result, the 
existence of AIC and TIC are theoretically guaranteed. 
However, in singular models, the Fisher information matrix becomes degenerate and its 
inverse does not exist. 
This phenomenon, of course, will appear in the case of quantum statistics. 
In Section~\ref{sec:numDemo}, we will actually see how this impact the model selection 
in practice.

We also note that there are other types of information criteria derived from other 
theoretical backgrounds, such as the Bayesian information criterion (BIC) 
\cite{schwarz1978Estimating}, which is based on the posterior probability distribution 
of models, and the minimum description length (MDL) \cite{rissanen1998Stochastic}, which 
is designed to minimize the description of the data. 
As in the classical statistics, there might be possibilities that one can derive other 
types of quantum information criteria than the ones derived in this paper, but here 
we focus on AIC and TIC and extend them to quantum statistics reflecting the 
correspondence between KL divergence and the quantum relative entropy.

Finally, the KL divergence defined in Eq. (\ref{eq:KL}) is not symmetric, and exchanging the role of the two arguments can make significant differences in the model selection. 
Ref. \cite{seghouane2004Small} indicates that the directed KL divergence $KL(g \|f)$ may better reflect the error due to overfitting, whereas the alternative directed divergence $KL(f\|g)$ may better reflect the error due to underfitting.
In Ref. \cite{seghouane2007AIC}, the authors addressed this issue and derived a new information criterion that is an asymptotically unbiased estimator of the symmetrized KL divergence. 
The quantum version of this proposal, based on the symmetrized quantum relative entropy, is worth investigating as a future work.

%%%%%%%%%%%%%%%%%%%%%%%%%%%%%%%%%%%%%%%%%%%%%%%%%%
%%%%%%%%%%%%%%%%%%%%%%%%%%%%%%%%%%%%%%%%%%%%%%%%%%
%%%%%%%%%%%%%%%%%%%%%%%%%%%%%%%%%%%%%%%%%%%%%%%%%%
\section{Quantum information criteria}
\label{sec:quantumInformationCriteria}

In this work, we develop the quantum version of the KL-based information criteria, AIC and TIC described in Section~\ref{sec:InformationCriteria}, by particularly 
replacing the KL divergence with the quantum relative entropy~\cite{umegaki1962Conditional}.
That is, we evaluate the quantum relative entropy \cite{wilde2017Quantum} between a true quantum state $\rho$ 
and a parametric quantum state $\{ \sigma(\bm\theta) : \bm\theta \in \Theta \subset \mathbb{R}^p \}$; 
\begin{equation}\label{eq:qre}
  D(\rho\|\sigma(\hat{\bm\theta})) 
    = \Tr \left[ \rho \left( \log \rho - \log \sigma(\hat{\bm\theta}) \right) \right].
\end{equation}
Here, we implicitly assume $\mathrm{supp}(\rho) \subseteq \mathrm{supp} (\sigma(\hat{\bm\theta}))$. 
In this paper, we further assume the regularity on $\sigma(\hat{\bm\theta})$, i.e., $\sigma(\hat{\bm\theta})$ is invertible.
$\hat{\bm\theta}$ is designed so that $\sigma(\hat{\bm\theta})$ well approximates 
$\rho$. 
Since the first term of quantum relative entropy does not depend on the model 
$\sigma(\hat{\bm\theta})$, it is enough to evaluate the second term for our purpose; 
in this paper, we call the second the quantum cross entropy (QCE).
In particular, following the idea behind the classical information criteria, we study 
the negative QCE:
\begin{equation}
\label{eq:negativeQCE}
  S(\rho\|\sigma(\hat{\bm\theta})) = \Tr \left( \rho \log \sigma(\hat{\bm\theta}) \right),
\end{equation}
which corresponds to Eq.~(\ref{eq:negativeCCE}). 
Our goal is to derive a good estimator for Eq.~(\ref{eq:negativeQCE}) and define 
the quantum information criterion. 
However, unlike the classical case where the log-likelihood function 
(\ref{eq:est_expected_loglikelihood}) appears as a natural estimator, it is unclear 
how to formulate a good estimator of the negative QCE. 
In this work, we present two variants, the one using the log-likelihood and the other 
using the classical shadow \cite{huang2020Predicting}. 
The resultant quantum information criterion using the log-likelihood, 
$\mathrm{QAIC}_{\mathrm{LL}}$, will be given in Eq.~(\ref{eq:QAIC_LL}); 
the quantum information criteria using the classical shadow, 
$\mathrm{QTIC}_{\mathrm{shadow}}$ and $\mathrm{QAIC}_{\mathrm{shadow}}$, will be 
given in Eq.~(\ref{eq:QTIC_shadow}) and Eq.~(\ref{eq:QAIC_shadow}), respectively. 
To the best of our knowledge, they are the first information criteria that are based on quantum relative entropy.
We note that $\mathrm{QAIC}_{\mathrm{LL}}$ uses the classical log-likelihood, but correcting the bias produces the quantum information-theoretic quantity.
Our derivation for these three information criteria is based on Ref.~\cite{konishi2008information}; the details of calculation for the bias 
is given in \ref{app:derivation}.
The asymptotic normality about estimators of $\bm\theta$, which is central to our derivation, is discussed in \ref{app:asymptotic_normality}.

%%%%%%%%%%%%%%%%%%%%%%%%%%%%%%
\subsection{Estimator of the negative QCE: Log-likelihood approach}
\label{subsec:qic_ll}

The first estimator of the negative QCE~(\ref{eq:negativeQCE}) is the averaged 
log-likelihood defined for a classical model distribution with fixed measurement; 
that is, roughly speaking, we use the classical cross entropy to estimate the 
quantum cross entropy.

To define an estimator, let us assume that we perform a tomographically complete measurement 
$\{\Pi_m\}$ on the true state $\rho$ and the measurement outcomes 
$\bm{x}_n = \{x_1, x_2, ..., x_n\}$ are sampled from the true probability mass function 
$g(x) = \Tr \left( \Pi_{x} \rho \right)$. 
Because the number of elements in $\{\Pi_m\}$ is assumed to be finite, $x$ is a discrete 
random variable. 
The tomographic completeness guarantees that, in general, every parameter in a parametric 
quantum state is asymptotically determined; 
conversely, if the measurement is not tomographically complete, some parameters might 
be impossible to estimate
\footnote{This argument assumes a general parameterization on quantum states. If observables to be measured for estimating parameters are obvious, in the case of QBM for instance, the measurement does not have to be tomographically complete.}. 
As in the classical case, we estimate the parameters using the maximum likelihood 
$\hat{\bm\theta}_C$~\footnote{The subscript $C$ represents "classical" to distinguish 
the maximum likelihood estimator $\hat{\bm\theta}_C$ from an estimator $\hat{\bm\theta}_Q$ 
($Q$ represents "quantum"), which will appear in the next subsection of the classical 
shadow's case. 
These subscripts represent if the cost function is classical or quantum. 
For example, $\hat{\bm\theta}_C$ is the maximizer of $\ell_{\mathrm{LL}}$, which is 
a consistent estimator of the negative classical cross entropy.}:
\begin{equation}
\label{eq:theta_C}
  \hat{\bm\theta}_C = \argmax_{\bm\theta \in \Theta} \ell_{\mathrm{LL}}(\bm{x}_n; \bm\theta), \quad \ell_{\mathrm{LL}}(\bm{x}_n; \bm\theta) = \sum_{\alpha=1}^n \log \Tr \left( \Pi_{x_\alpha} \sigma(\bm\theta) \right).
\end{equation}
Then we use the log-likelihood 
\begin{equation}
\label{eq:est_negativeQCE}
  \mathbb{E}_{\hat{g}(z)} \left[ \log \Tr \left( \Pi_{Z} \sigma(\hat{\bm\theta}_C) \right) \right] = \frac{1}{n} \ell_{\mathrm{LL}}(\bm{x}_n; \hat{\bm\theta}_C)
\end{equation}
to estimate the negative QCE, where $\hat{g}(z)$ is the empirical probability mass 
function. 
As in the classical case, this is a biased estimator in the sense that the expected 
log-likelihood $\mathbb{E}_{g(z)} \left[ \log \Tr \left( \Pi_{Z} \sigma(\hat{\bm\theta}_C) \right) \right]$ is not equal to the negative QCE, and the bias-corrected estimator is 
our target quantum information criterion. 
Now let $\bm\theta_0$ be a solution to $\Tr \left( \rho \frac{\partial \log \sigma(\bm\theta)}{\partial \bm\theta} \right) = 0$. 
Then, despite the bias, the maximum likelihood estimator $\hat{\bm\theta}_C$ converges 
to $\bm\theta_0$ when $n \rightarrow \infty$; see \ref{app:proof_of_convergence} for 
the proof. 
This is the main motivation to use the log-likelihood $\ell_{\mathrm{LL}}(\bm{x}_n; \hat{\bm\theta}_C)$.

The bias of the estimator using the log-likelihood is defined as follows:
\begin{eqnarray}
    b(G) &= \mathbb{E}_{g(\bm x_n)} \left[ \ell_{\mathrm{LL}}(\bm X_n ; \hat{\bm\theta}_C) - n \Tr \left( \rho \log \sigma(\hat{\bm\theta}_C) \right) \right] \nonumber\\
    &= \mathbb{E}_{g(\bm x_n)} \left[ \ell_{\mathrm{LL}}(\bm X_n ; \hat{\bm\theta}_C) - \ell_{\mathrm{LL}}(\bm X_n ; \bm\theta_0) \right] \nonumber\\
    & \quad \, + \mathbb{E}_{g(\bm x_n)} \left[ \ell_{\mathrm{LL}}(\bm X_n ; \bm\theta_0) - n \Tr \left( \rho \log \sigma(\bm\theta_0) \right) \right]\nonumber\\
    & \quad \, + \mathbb{E}_{g(\bm x_n)} \left[ n \Tr \left( \rho \log \sigma(\bm\theta_0) \right) - n \Tr \left( \rho \log \sigma(\hat{\bm\theta}_C) \right) \right] \nonumber\\
    &= D_1 + D_2 + D_3, \nonumber
\end{eqnarray}
where the expectation $\mathbb{E}_{g(\bm{x}_n)}$ is taken with respect to the joint 
probability mass function $g(\bm{x}_n) = \prod_{\alpha=1}^{n} g(x_\alpha)$. 
To derive an unbiased estimator for the negative QCE, we give an explicit form of the bias. 
The explicit calculations of $D_1$, $D_2$, and $D_3$, 
which are detailed in \ref{app:derivation_LL},
conclude that the bias of the estimation of the negative 
QCE using the log-likelihood is obtained, in the asymptotic regime, as
\begin{eqnarray}
  b(G) &= D_1 + D_2 + D_3 \nonumber\\
  &= \frac{1}{2} \Tr \left( I_C(\bm\theta_0) J_C(\bm\theta_0)^{-1} \right) \nonumber\\
  & \quad \, + n \left( H(g(z)\|\Tr \left( \Pi_{z} \sigma(\bm\theta_0) \right)) - \Tr \left( \rho \log \sigma(\bm\theta_0) \right) \right) \nonumber\\
  & \quad \, + \frac{1}{2} \Tr \left( J_Q(\bm\theta_0) J_C(\bm\theta_0)^{-1} I_C(\bm\theta_0) J_C(\bm\theta_0)^{-1} \right), \label{eq:LL_bias}
\end{eqnarray}
where $J_Q(\bm\theta_0)$, $I_C(\bm\theta_0)$, and $J_C(\bm\theta_0)$ are defined in Eqs. (\ref{eq:J_Q}), (\ref{eq:LL_I_C}), and (\ref{eq:LL_J_C}), respectively.
The second term of the bias~(\ref{eq:LL_bias}) cannot be computed in practice, 
because it includes the classical and quantum cross entropy between the true 
state and the optimal parametric state. 
This prevents us from defining the quantum version of TIC (\ref{eq:TIC}) using the log-likelihood.
To avoid this issue, we assume that the parametric quantum state 
$\{ \sigma(\bm\theta) : \bm\theta \in \Theta \subset \mathbb{R}^p \}$ is realizable, 
meaning that there exists $\bm\theta_0$ satisfying $\rho = \sigma(\bm\theta_0)$. 
In this case, using the relation $J_C(\bm\theta_0) = I_C(\bm\theta_0)$, the above 
bias expression is reduced to
\begin{eqnarray}
  \fl b(G) &= D_1 + D_2 + D_3 \nonumber\\
  \fl &= \frac{1}{2} p + n \left( H(g(z)\|g(z)) - \Tr \left( \rho \log \rho \right) \right) + \frac{1}{2} \Tr \left( J_Q(\bm\theta_0) I_C(\bm\theta_0)^{-1} \right). 
\label{eq:LL_bias_reduced}
\end{eqnarray}
The second term of the above reduced bias is independent of the model, and thus 
it is not necessary to compute when comparing the values of different models. 
Now note that $J_Q(\bm\theta_0)$ and $I_C(\bm\theta_0)$, which are defined in 
Eqs. (\ref{eq:J_Q}) and (\ref{eq:LL_I_C}) respectively, contain the unknown 
factors $\rho$ and $g(x) = \Tr \left( \Pi_{x} \rho \right)$. 
This issue can also be resolved by the realizability assumption, which allows us 
to replace them by $\sigma(\bm\theta_0)$ and $g(x) = \Tr \left( \Pi_{x} \sigma(\bm\theta_0) \right)$, respectively. 
As a result, $J_Q(\bm\theta_0)$ and $I_C(\bm\theta_0)$ can be replaced with 
their consistent estimators 
\begin{eqnarray}
  \hat{J}_{Q;ij}(\hat{\bm\theta}_C) &= - \Tr \left( \sigma(\hat{\bm\theta}_C) \left. \frac{\partial^2 \log \sigma(\bm\theta)}{\partial \theta_i \partial \theta_j} \right|_{\bm\theta = \hat{\bm\theta}_C} \right), \label{eq:LL_J_Q_est} \\
  \hat{I}_{C;ij}(\hat{\bm\theta}_C) &= \mathbb{E}_{h(x)} \left[ \left. \frac{\partial \log \Tr \left( \Pi_X \sigma(\bm\theta) \right) }{\partial \theta_i} \frac{\partial \log \Tr \left( \Pi_X \sigma(\bm\theta) \right)}{\partial \theta_j} \right|_{\bm\theta = \hat{\bm\theta}_C} \right], 
\label{eq:LL_I_C_est}
\end{eqnarray}
where $\mathbb{E}_{h(x)}$ is the expectation with respect to the probability mass 
function $h(x) = \Tr(\Pi_x \sigma(\hat{\bm\theta}_C))$. 
Importantly, $\hat{J}_Q(\hat{\bm\theta}_C)$ is the Bogoljubov Fisher information matrix 
at $\hat{\bm\theta}_C$~\cite{hasegawa1997Exponential,amari2000methods,hayashi2002Two}. 
We give a brief review of quantum Fisher information, including Bogoljubov Fisher information, in \ref{app:quntumFisherInfo}.
From Eqs.~(\ref{eq:LL_bias_reduced}),(\ref{eq:LL_J_Q_est}), and (\ref{eq:LL_I_C_est}), 
we define the quantum information criterion using the log-likelihood: 
\begin{equation}\label{eq:QAIC_LL}
  \mathrm{QAIC}_{\mathrm{LL}} 
     = - 2 \ell_{\mathrm{LL}}(\hat{\bm\theta}_C) + p 
          + \Tr \left( \hat{J}_Q(\hat{\bm\theta}_C) \hat{I}_C(\hat{\bm\theta}_C)^{-1} \right).
\end{equation}
Here we omit the second term of Eq.~(\ref{eq:LL_bias_reduced}), as it is 
a model-independent constant that is not necessary for comparing different 
parametric models. 
Note that $-\mathrm{QAIC}_{\mathrm{LL}}/(2n)$ with this constant asymptotically becomes an unbiased estimator of the negative QCE:
\begin{eqnarray*}
  \mathbb{E}_{g(\bm{x}_n)}\left[ - \frac{1}{2n} \mathrm{QAIC}_{\mathrm{LL}} - \left( H(g(z)\|g(z)) - \Tr \left( \rho \log \rho \right) \right) \right] \\
  \quad = \mathbb{E}_{g(\bm{x}_n)}\left[ \Tr \left( \rho \log \sigma(\hat{\bm\theta}_C) \right) \right] + o\left( n^{-1} \right).
\end{eqnarray*}

We recall that the quantum Fisher information matrix $\hat{J}_Q$ is equal to or greater than the 
classical Fisher information matrix $\hat{I}_C$, i.e., $\hat{J}_Q \geq \hat{I}_C$, with equality if and only 
if the optimal measurement strategy is performed for estimating the parameters $\bm\theta$.
This concludes that the third term of $\mathrm{QAIC}_{\mathrm{LL}}$ (\ref{eq:QAIC_LL}) is 
always equal to or greater than $p$, meaning $\mathrm{QAIC}_{\mathrm{LL}} \geq \mathrm{AIC}$ 
in general. 
Particularly it is interesting to see the case of $\hat{J}_Q = \hat{I}_C$, that leads to 
$\mathrm{QAIC}_{\mathrm{LL}} = - 2 \ell_{\mathrm{LL}}(\hat{\bm\theta}_C) + 2p = \mathrm{AIC}$. 
Now, $\hat{J}_Q = \hat{I}_C$ is achieved only when observables associated with parameters can be 
simultaneously measured and the corresponding projective measurement is chosen.
This is exactly equivalent to the classical scenario, where the computational basis 
measurement is performed to estimate parameters that are associated with commuting 
observables. 
This result coincides with the intuition that the bias of the estimator should be larger 
if the measurement is not optimal, because the bias originates from the fluctuation from 
the parameter estimation.
This kind of measurement-induced bias, which does not exist in the classical case, should 
be incorporated in model selection because the optimality of the chosen measurement is 
different among parametric models.

%%%%%%%%%%%%%%%%%%%%%%%%%%%%%%
\subsection{Estimator of the negative QCE: Classical shadow approach}
\label{subsec:qic_shadow}

The second estimator of the negative QCE~(\ref{eq:negativeQCE}) is the average of 
negative QCEs defined for a classical shadow state; 
that is, roughly speaking, we use the quantum cross entropy with classical 
shadow, i.e., ``quantum log-likelihood", to estimate the quantum cross entropy.

The classical shadow \cite{huang2020Predicting} can be used to efficiently estimate 
observables with a finite number of measurements. 
In our case, instead of the outcomes $\bm{x}_n = \{x_1, x_2, ..., x_n\}$ obtained 
by measuring an unknown true state $\rho$, we use the classical shadow 
$\hat{\bm\rho}_n = \{\hat{\rho}_1, \hat{\rho}_2, ..., \hat{\rho}_n\}$ to 
construct an estimator for the negative QCE. 
Originally, to construct the classical shadow, we repeatedly apply a random unitary 
$U$ sampled from the set $\mathcal{U}$ to rotate the state and then perform 
a computational-basis measurement to get the measurement outcome 
$\ket{\hat{b}}$, $\hat{b} \in \Omega$, where $\Omega$ is a computational basis domain. 
Here we assume that the set of unitaries $\mathcal{U}$ defines a tomographically 
complete set of measurements so that the inverse of the quantum channel 
$\mathcal{M}^{-1}$ exists.
After the measurement, we apply the inverse of $U$ and $\mathcal{M}^{-1}$ to get 
a classical {\it snapshot} $\hat{\rho} = \mathcal{M}^{-1} \left( U^\dagger \ket{\hat{b}} \bra{\hat{b}} U \right)$. 
Thus, $\hat{\rho}$ is sampled with probability 
$g(\hat{\rho}) = \bra{\hat{b}} U \rho U^\dagger \ket{\hat{b}}$, by definition.
This snapshot has a preferable feature for our derivation; it reconstructs the true 
state in expectation, $\mathbb{E}[\hat{\rho}] = \rho$, and has statistical convergence guarantees \cite{guta2020Fast}. 
Based on this property, we define 
\begin{equation}
   \hat{\bm\theta}_Q 
      = \argmax_{\bm\theta \in \Theta} 
            \ell_{\mathrm{shadow}}(\hat{\bm\rho}_n; \bm\theta), ~~~
  \ell_{\mathrm{shadow}}(\hat{\bm\rho}_n; \bm\theta) 
      = \sum_{\alpha=1}^n 
          \Tr \left( \hat{\rho}_\alpha \log \sigma(\bm\theta) \right), 
\end{equation}
and the estimator for the negative QCE, $\Tr( \rho \log \sigma(\hat{\bm\theta}_Q) )$, as
\begin{equation}
\label{eq:est_negativeQCE_shadow}
  \mathbb{E}_{\hat{g}(\hat{\rho})} 
     \left[ \Tr \left( \hat{\rho} \log \sigma(\hat{\bm\theta}_Q) \right) \right] 
        = \frac{1}{n} \ell_{\mathrm{shadow}}(\hat{\bm\rho}_n; \hat{\bm\theta}_Q),
\end{equation}
where $\hat{g}(\hat{\rho})$ is the empirical probability mass function. 
However, because Eq.~(\ref{eq:est_negativeQCE_shadow}) contains $\hat{\bm\theta}_Q$ 
which is a function of the snapshots, it is not an unbiased estimator for the negative 
QCE. 

The bias of the estimator using the classical shadow is defined as follows:
\begin{eqnarray*}
  b(G) &= \mathbb{E}_{g(\hat{\bm\rho}_n)} \left[ \ell_{\mathrm{shadow}}(\hat{\bm\rho}_n; \hat{\bm\theta}_Q) - n \Tr \left( \rho \log \sigma(\hat{\bm\theta}_Q) \right) \right] \\
  &= \mathbb{E}_{g(\hat{\bm\rho}_n)} \left[ \ell_{\mathrm{shadow}}(\hat{\bm\rho}_n; \hat{\bm\theta}_Q) - \ell_{\mathrm{shadow}}(\hat{\bm\rho}_n; \hat{\bm\theta}_0) \right] \\
  & \quad \, + \mathbb{E}_{g(\hat{\bm\rho}_n)} \left[ \ell_{\mathrm{shadow}}(\hat{\bm\rho}_n; \hat{\bm\theta}_0) - n \Tr \left( \rho \log \sigma(\bm\theta_0) \right) \right] \\
  & \quad \, + \mathbb{E}_{g(\hat{\bm\rho}_n)} \left[ n \Tr \left( \rho \log \sigma(\bm\theta_0) \right) - n \Tr \left( \rho \log \sigma(\hat{\bm\theta}_Q) \right) \right] \\
  &= D_1 + D_2 + D_3,
\end{eqnarray*}
where the expectation $\mathbb{E}_{g(\hat{\bm\rho}_n)}$ is taken with respect to the 
joint probability density function, 
$g(\hat{\bm\rho}_n) = \prod_{\alpha=1}^n g(\hat{\rho}_\alpha)$. 
Also, $\bm\theta_0$ is the solution to $\Tr \left( \rho \frac{\partial \log \sigma(\bm\theta)}{\partial \bm\theta} \right) = 0$. 
To derive an unbiased estimator for the negative QCE, we give an explicit form of the bias.
The explicit calculations of $D_1$, $D_2$, and $D_3$, 
which are detailed in \ref{app:derivation_shadow},
conclude that the bias of the shadow-based estimator of the 
negative QCE is obtained, in the asymptotic regime, as 
\begin{eqnarray}
  b(G) &= D_1 + D_2 + D_3 \nonumber\\
  &= \frac{1}{2} \Tr \left( I_Q(\bm\theta_0) J_Q(\bm\theta_0)^{-1} \right) + 0 + \frac{1}{2} \Tr \left( I_Q(\bm\theta_0) J_Q(\bm\theta_0)^{-1} \right) \nonumber\\
  &= \Tr \left( I_Q(\bm\theta_0) J_Q(\bm\theta_0)^{-1} \right). 
\label{eq:shadow_bias}
\end{eqnarray}
Note now that $I_Q(\bm\theta_0)$ and $J_Q(\bm\theta_0)$, 
defined in Eqs. (\ref{eq:I_Q_shadow}) and (\ref{eq:J_Q}) respectively, depend on the unknown distribution 
$g(\hat{\rho})$ and state $\rho$; we thus replace these terms with their consistent estimators
\begin{eqnarray}\label{eq:shadow_I_Q_est}
  \hat{I}^{\mathrm{emp}}_{Q;ij}(\hat{\bm\theta}_Q) &= \mathbb{E}_{\hat{g}(\hat{\rho})} \left[ \Tr \left( \hat{\rho} \left. \frac{\partial \log \sigma(\bm\theta)}{\partial \theta_i} \right|_{\bm\theta = \bm\theta_Q} \right) \Tr \left( \hat{\rho} \left. \frac{\partial \log \sigma(\bm\theta)}{\partial \theta_j} \right|_{\bm\theta = \bm\theta_Q} \right) \right] \nonumber\\
  &= \frac{1}{n} \sum_{\alpha=1}^{n} \Tr \left( \hat\rho_\alpha \left. \frac{\partial \log \sigma(\bm\theta)}{\partial \theta_i} \right|_{\bm\theta = \hat{\bm\theta}_Q} \right) \Tr \left( \hat\rho_\alpha \left. \frac{\partial \log \sigma(\bm\theta)}{\partial \theta_j} \right|_{\bm\theta = \hat{\bm\theta}_Q} \right),
\end{eqnarray}
\begin{eqnarray}\label{eq:shadow_J_Q_est}
  \hat{J}^{\mathrm{emp}}_{Q;ij}(\hat{\bm\theta}_Q) &= \mathbb{E}_{\hat{g}(\hat{\rho})} \left[ - \Tr \left( \hat{\rho} \left. \frac{\partial^2 \log \sigma(\bm\theta)}{\partial \theta_i \partial \theta_j} \right|_{\bm\theta = \hat{\bm\theta}_Q} \right) \right] \nonumber\\
  &= - \Tr \left( \left( \frac{1}{n} \sum_{\alpha=1}^n \hat{\rho}_\alpha \right) \left. \frac{\partial^2 \log \sigma(\bm\theta)}{\partial \theta_i \partial \theta_j} \right|_{\bm\theta = \hat{\bm\theta}_Q} \right).
\end{eqnarray}
Using Eqs.~(\ref{eq:shadow_bias}), (\ref{eq:shadow_I_Q_est}), and (\ref{eq:shadow_J_Q_est}), 
we define the quantum information criterion based on the classical shadow, 
which is a quantum version of TIC (\ref{eq:TIC}): 
\begin{equation}\label{eq:QTIC_shadow}
  \mathrm{QTIC}_{\mathrm{shadow}} = - 2 \ell_{\mathrm{shadow}}(\hat{\bm\rho}_n ; \hat{\bm\theta}_Q) + 2 \Tr \left( \hat{I}^{\mathrm{emp}}_Q(\hat{\bm\theta}_Q) \hat{J}^{\mathrm{emp}}_Q(\hat{\bm\theta}_Q)^{-1} \right).
\end{equation}

The realizability assumption that the set of parametric quantum states includes the 
true state allows us to simplify $\mathrm{QTIC}_{\mathrm{shadow}}$. 
Actually, under this assumption, $I_Q(\bm\theta_0)$ and $J_Q(\bm\theta_0)$ 
appearing in Eq.~(\ref{eq:shadow_bias}) are replaced with
\begin{equation}
\label{eq:shadow_hat_I_Q}
  \hat{I}_{Q;ij}(\hat{\bm\theta}_Q) = \mathbb{E}_{h(\hat{\rho})} \left[ \Tr \left( \hat{\rho} \left. \frac{\partial \log \sigma(\bm\theta)}{\partial \theta_i} \right|_{\bm\theta = \hat{\bm\theta}_Q}  \right) \Tr \left( \hat{\rho} \left. \frac{\partial \log \sigma(\bm\theta)}{\partial \theta_j} \right|_{\bm\theta = \hat{\bm\theta}_Q}  \right) \right],
\end{equation}
\begin{equation}
\label{eq:shadow_hat_J_Q}
  \hat{J}_{Q;ij}(\hat{\bm\theta}_Q) = - \Tr \left( \sigma(\hat{\bm\theta}_Q) \left. \frac{\partial^2 \log \sigma(\bm\theta)}{\partial \theta_i \partial \theta_j} \right|_{\bm\theta = \hat{\bm\theta}_Q} \right),
\end{equation}
where $h(\hat{\rho}) = \bra{\hat{b}} U \sigma(\hat{\bm\theta}_Q) U^\dagger \ket{\hat{b}}$, 
and thus we obtain 
\begin{equation}
\label{eq:QAIC_shadow}
  \mathrm{QAIC}_{\mathrm{shadow}} = - 2 \ell_{\mathrm{shadow}}(\hat{\bm\rho}_n ; \hat{\bm\theta}_Q) + 2 \Tr \left( \hat{I}_Q(\hat{\bm\theta}_Q) \hat{J}_Q(\hat{\bm\theta}_Q)^{-1} \right).
\end{equation}
Importantly, unlike the conventional $\mathrm{AIC}$, the second term of 
$\mathrm{QAIC}_{\mathrm{shadow}}$ is not equal to the number of parameters. 
Note also that $\mathrm{QTIC}_{\mathrm{shadow}}$ and $\mathrm{QAIC}_{\mathrm{shadow}}$ 
multiplied by $-1/(2n)$ are asymptotic unbiased estimators of the negative QCE: 
\begin{eqnarray*}
  \mathbb{E}_{g(\hat{\bm\rho}_n)}\left[ - \frac{1}{2n} \mathrm{QTIC}_{\mathrm{shadow}} \right] = \mathbb{E}_{g(\hat{\bm\rho}_n)}\left[ \Tr \left( \rho \log \sigma(\hat{\bm\theta}_Q) \right) \right] + o\left( n^{-1} \right), \\
  \mathbb{E}_{g(\hat{\bm\rho}_n)}\left[ - \frac{1}{2n} \mathrm{QAIC}_{\mathrm{shadow}} \right] = \mathbb{E}_{g(\hat{\bm\rho}_n)}\left[ \Tr \left( \rho \log \sigma(\hat{\bm\theta}_Q) \right) \right] + o\left( n^{-1} \right).
\end{eqnarray*}
\vspace{5mm}

For the rest of this section, we discuss how to compute $\mathrm{QTIC}_{\mathrm{shadow}}$ 
or $\mathrm{QAIC}_{\mathrm{shadow}}$. 
As demonstrated above, though the classical shadow is useful for deriving these criteria, 
it was originally designed to approximate quasi-local observables. 
In general, $\log \sigma(\bm\theta)$ is a global observable and thus we need to carefully 
choose a parametric quantum state $\sigma(\bm\theta)$, because otherwise the estimation 
might have large variance. 
In addition, computation of the matrix logarithm $\log \sigma(\bm\theta)$ and its 
derivatives cannot be efficiently conducted via classical means. 
To mitigate these issues, we choose the quantum model $\sigma(\bm\theta)$ from a class 
of quantum Boltzmann machines (QBM). 
In particular, a thermal state generated by QBM is represented as
\begin{equation*}
  \sigma_{\mathrm{QBM}}(\bm\theta) = \frac{\exp(- H_{\bm\theta} / (k_B T))}{Z_{\bm\theta}}, \quad Z_{\bm\theta} = \Tr \left( \exp(- H_{\bm\theta} / (k_B T)) \right),
\end{equation*}
with $k_B$ the Boltzmann constant, $T$ the system temperature, and $Z_{\bm\theta}$ the 
partition function.
Here we define a parameterized Hamiltonian $H_{\bm\theta} = \sum_{i=0}^{p-1} \theta_i P_i$ 
where $P_i = \otimes_j \sigma_{j,i}$ with $\sigma_{j,i} \in \{I, X, Y, Z\}$ the Pauli 
operator acting on the $j$-th qubit. 
Assuming that $k_B T = 1$ for simplicity, we easily see that
\begin{equation*}
    \log \sigma_{\mathrm{QBM}}(\bm\theta) = - H_{\bm\theta} - \log Z_{\bm\theta}.
\end{equation*}
Thus, the quasi-locality of $\log \sigma_{\mathrm{QBM}}(\bm\theta)$ can be satisfied by 
choosing appropriate Pauli strings in $H_{\bm\theta}$. 
Also, the computational hardness of the matrix logarithm is now converted to the hardness 
of the log-partition function $\log Z_{\bm\theta}$. 
Moreover, Ref.~\cite{verdon2019Quantum} proposed a quantum model called the quantum 
Hamiltonian-based models (QHBMs) that have an explicit representation of the exponential 
form and can be implemented on a quantum-classical hybrid computer. 
QHBMs can turn the quantum log-partition function into the classical one, which actually 
mitigates the computational hardness of the partition function. 
We will use QHBM in the numerical demonstration of Section~\ref{sec:numDemo}; 
it will be shown there that we can obtain the analytic form of terms like 
$\partial_i \log \sigma_{\mathrm{QHBM}}(\bm\theta)$ and 
$\partial_i \partial_j \log \sigma_{\mathrm{QHBM}}(\bm\theta)$, as detailed in 
Ref.~\cite{verdon2019Quantum,sbahi2022Provably}.

Other than the above-mentioned approach, recently a combination of structural assumptions 
on quantum states with the classical shadow has been found to achieve better sample 
complexity for estimating observables. 
It has been shown that exponential improvements in the sample complexity hold for 
some specific quantum states such as high-temperature Gibbs states of commuting
Hamiltonians or outputs of shallow circuits \cite{rouze2021Learning}; quantum Gibbs 
states of non-commuting Hamiltonians with exponential decay of correlations 
\cite{onorati2023Efficient}; ground states of gapped local Hamiltonians 
\cite{lewis2023Improved}. 
Thus, there is another room for improvements to reduce the variance of the estimation 
in practice.

%%%%%%%%%%%%%%%%%%%%%%%%%%%%%%%%%%%%%%%%%%%%%%%%%%
\section{Numerical demonstration}
\label{sec:numDemo}

In this section, we numerically study a quantum state estimation problem using a classical backend, instead of a quantum processing unit, and evaluate the 
performance of the quantum information criteria for model selection. 
Among the three quantum information criteria derived in Section 
\ref{sec:quantumInformationCriteria}, we use $\mathrm{QTIC}_{\mathrm{shadow}}$ because 
the analytical form of the first and second derivatives of $\ell_{\mathrm{shadow}}$ can 
be obtained for the problem to be investigated. 
Note that $\mathrm{QAIC}_{\mathrm{shadow}}$ is also calculable, but the matrix $\hat{I}_{Q}$ 
given in Eq.~(\ref{eq:shadow_hat_I_Q}) takes the expectation with respect to every possible 
classical snapshot, whose total number is much bigger than the number of samples $n$ for 
large systems. 
In addition, $\mathrm{QAIC}_{\mathrm{shadow}}$ assumes that the set of parametric quantum 
states includes a true state, while $\mathrm{QTIC}_{\mathrm{shadow}}$ does not need this 
assumption. 
From here, we simply refer $\mathrm{QTIC}_{\mathrm{shadow}}$ as $\mathrm{QTIC}$. 
The goal is to test whether $\mathrm{QTIC}$ would properly work or not, by comparing it 
with the negative QCE between the true quantum state $\rho$ and the estimated quantum state 
$\sigma(\hat{\bm\theta})$, i.e., 
\begin{equation*}
  \mathrm{QCE}_\mathrm{true} = - \Tr \left( \rho \log \sigma(\hat{\bm\theta}) \right),
\end{equation*}
which cannot be calculated in practice because the true state is unknown. 
We note that the value of $\mathrm{QTIC}$ is divided by $1/(2n)$ to compare with 
$\mathrm{QCE}_\mathrm{true}$ in the following numerical results.

To estimate the true state, we use QHBM~\cite{verdon2019Quantum} as a parametric quantum 
state $\sigma(\bm\theta)$. 
QHBM is a quantum analogue of a classical energy-based model, consisting of classical and 
quantum parts. 
The classical part is an energy-based model (EBM) represented by the energy function 
$E_{\bm\theta'}(\bm{x})$ with latent variational parameters $\bm\theta'$ and a binary 
string $\bm{x}$. 
The quantum part is a parameterized quantum circuit (PQC) represented by the unitary 
operator $U(\bm\theta'')$ with model parameters $\bm\theta''$. 
The variational mixed state is then
\begin{equation*}
  \sigma_{\mathrm{QHBM}}(\bm\theta) 
   = \frac{1}{Z_{\bm\theta'}} \exp(- U(\bm\theta'') K_{\bm\theta'} U^\dagger(\bm\theta'')),
  \quad \bm\theta = (\bm\theta', \bm\theta''),
\end{equation*}
where $K_{\bm\theta'}$ is called the latent modular Hamiltonian given by 
\begin{equation*}
    K_{\bm\theta'} 
      = \sum_{\bm{x} \in \Omega} E_{\bm\theta'}(\bm{x}) \ket{\bm{x}}\bra{\bm{x}},
\end{equation*}
with $\Omega$ the computational basis domain. 
Also $Z_{\bm\theta'} = \Tr(\exp(- K_{\bm\theta'}))$ is the model partition function. 
In the numerical simulation, we use the fully-connected Boltzmann machine as EBM.

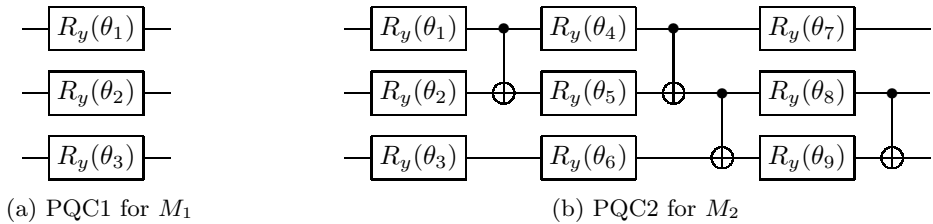
\begin{figure}[tb]
  \centering
  \begin{subfigure}[b]{0.25\textwidth}
      \mbox{
      \Qcircuit @C=1.0em @R=0.8em {
          & \push{\rule{1em}{0em}} & \gate{R_y(\theta_1)} & \qw & \push{\rule{1em}{0em}} \\
          & \push{\rule{1em}{0em}} & \gate{R_y(\theta_2)} & \qw & \push{\rule{1em}{0em}} \\
          & \push{\rule{1em}{0em}} & \gate{R_y(\theta_3)} & \qw & \push{\rule{1em}{0em}} \\
      }
      }
      \caption{PQC1 for $M_1$}
      \label{fig:pqc1}
  \end{subfigure}
  \hfil
  \begin{subfigure}[b]{0.65\textwidth}
      \mbox{
      \Qcircuit @C=1.0em @R=0.8em {
          & \push{\rule{1em}{0em}} & \gate{R_y(\theta_1)} & \ctrl{1} & \gate{R_y(\theta_4)} & \ctrl{1} & \qw      & \gate{R_y(\theta_7)} & \qw      & \qw \\
          & \push{\rule{1em}{0em}} & \gate{R_y(\theta_2)} & \targ    & \gate{R_y(\theta_5)} & \targ    & \ctrl{1} & \gate{R_y(\theta_8)} & \ctrl{1} & \qw \\
          & \push{\rule{1em}{0em}} & \gate{R_y(\theta_3)} & \qw      & \gate{R_y(\theta_6)} & \qw      & \targ    & \gate{R_y(\theta_9)} & \targ    & \qw \\
      }
      }
      \caption{PQC2 for $M_2$}
      \label{fig:pqc2}
  \end{subfigure}
  \caption{PQCs for parametric quantum states $M_1$ and $M_2$.}
  \label{fig:pqc}
\end{figure}

For the demonstration of calculating $\mathrm{QTIC}$, we consider a $3$-qubit system 
and compare two parametric quantum states represented by QHBM against some fixed true 
state. 
To prepare two parametric quantum state models $M_1$ and $M_2$, we use two different 
PQCs shown by PQC1 and PQC2 in Fig.~\ref{fig:pqc}. 
PQC1 consists of a single layer of $R_y$ rotation gates, each of which is given by 
$R_y(\theta)={\rm exp}(-i\theta \sigma_y)$; hence PQC1 contains $3$ real parameters. 
PQC2 consists of three layers of $R_y$ gates and some CNOT gates, resulting in $9$ 
real parameters. 
As for the classical part, we use a standard fully-connected Boltzmann machine with 
three nodes. 
Since it has $6$ real variational parameters, the parametric quantum state models $M_1$ 
and $M_2$ have $9$ and $15$ parameters in total, respectively. 
Note that models $M_1$ and $M_2$ constitute a hierarchical series of models such 
that $M_1 \subset M_2$. 
We set the true state to be $M_1$ with randomly chosen parameters; note that the 
behavior of the value of $\mathrm{QTIC}$ depends on the true state, but we observed 
similar behavior for other true states as the case shown below.

Figure~\ref{fig:num_proc} shows the procedure of the numerical simulation.
In Step 1, we perform the random Pauli measurement on the true state.
The number of measurements, or equivalently the number of classical snapshots, is 
set to $n=1000$. 
In Step 2, the parameters are estimated based on the measurement outcome obtained 
in Step 1.
We use L-BFGS-B in SciPy~\cite{2020SciPy-NMeth} to minimize $- \ell_{\mathrm{shadow}}(\hat{\bm\rho}_n; \bm\theta)$ for the parameter estimation. 
In Step 3, $\mathrm{QTIC}$ is calculated based on the measurement outcome obtained 
in Step 1. 
Note that Steps 2 and 3 are conducted for both models $M_1$ and $M_2$.
We use the analytic form of the first and second derivatives of the logarithm 
of the density matrix, $\partial_i \log \sigma_{\mathrm{QHBM}}(\bm\theta)$ and 
$\partial_i \partial_j \log \sigma_{\mathrm{QHBM}}(\bm\theta)$, derived in 
Refs.~\cite{verdon2019Quantum,sbahi2022Provably}, for these steps.
We note that Refs.~\cite{verdon2019Quantum,sbahi2022Provably} proposed a method to 
compute the derivatives via the parameter shift rule on a quantum computer, but the 
derivatives are supposed to be computed on classical computers in our setting to 
ensure that all the estimates are based on the same measurement data of $n$ 
classical snapshots. 
This computation of derivative is consistent with the definition of the quantum information criteria introduced in
Section~\ref{sec:quantumInformationCriteria}.
In Step 4, the model selection is finally performed using the values of $\mathrm{QTIC}$ obtained in Step 3.
Furthermore, to investigate the effect of deviation caused by the finite number of 
measurements, we repeat the measurement and estimation procedures $50$ times for 
the same true state.
\begin{figure}[tb]
  \centering
  \includegraphics[width=\textwidth]{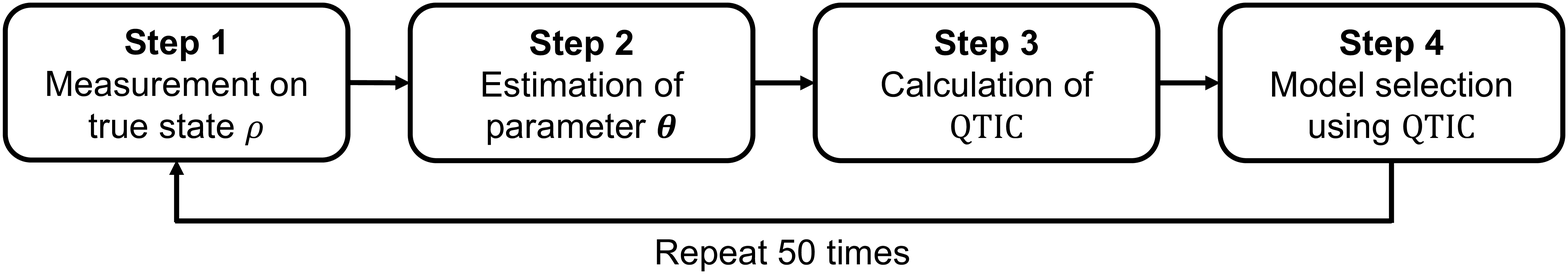}
  \caption{Procedure of the numerical simulation.}
  \label{fig:num_proc}
\end{figure}

\begin{figure}[tb]
  \centering
  \includegraphics[width=0.7\textwidth]{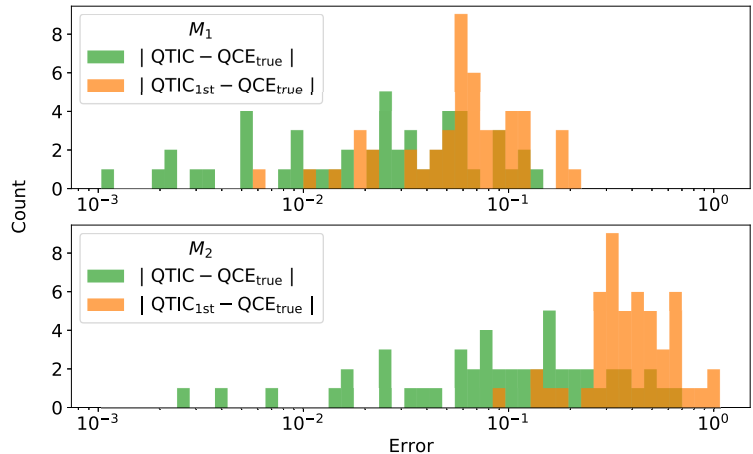}
  \caption{
  Error of $\mathrm{QTIC}$ against $\mathrm{QCE}_\mathrm{true}$ for the parametric 
  quantum states $M_1$ and $M_2$ (green).
  The horizontal line represents the value of error in the log plot and the vertical 
  line represents the number of counts out of 50 independent experiments. 
  The error of the $1^\mathrm{st}$ term of $\mathrm{QTIC}$, denoted as 
  $\mathrm{QTIC}_\mathrm{1st}$, against $\mathrm{QCE}_\mathrm{true}$ is shown (orange).}
  \label{fig:diff}
\end{figure}

Before showing the result of model selection, we compute the error of $\mathrm{QTIC}$ 
against $\mathrm{QCE}_\mathrm{true}$. 
Figure~\ref{fig:diff} shows the histogram of the values of $|\mathrm{QTIC}-\mathrm{QCE}_\mathrm{true}|$ for the parametric quantum states $M_1$ and $M_2$ (green). 
The horizontal line represents the value of error in the log plot and the vertical 
line represents the number of counts out of 50 independent experiments. 
For comparison, the error of the 1st term of $\mathrm{QTIC}$, which is 
denoted as $\mathrm{QTIC}_\mathrm{1st}$ (i.e., the loss function used for estimating 
the parameters) against $\mathrm{QCE}_\mathrm{true}$ is shown in the orange histogram. 
Comparing these two histograms, we find that correcting the bias by the second term 
of $\mathrm{QTIC}$ leads to smaller error, for both cases of $M_1$ and $M_2$. 
This clearly means that the bias-corrected $\mathrm{QTIC}$ works as a valid measure 
for model selection criterion in terms of the quantum relative entropy.

Next, we perform the model selection based on $\mathrm{QTIC}$.
Table~\ref{tab:ranking} shows the number of selected models out of 50 experiments, 
for the cases of using $\mathrm{QTIC}_\mathrm{1st}$ and $\mathrm{QTIC}$. 
For reference, the result when using $\mathrm{QCE}_{\mathrm{true}}$, i.e., the generalization error
$-\Tr[ \rho \log \sigma(\hat{\bm\theta}) ]$, is also shown; in this case, $M_1$ is 
always chosen because $M_1$ can realize the true state $\rho$ with less parameters than $M_2$. 
On the other hand, $\mathrm{QTIC}_\mathrm{1st}$ selects $M_2$ every time in 50 
experiments. 
This makes sense because $\mathrm{QTIC}_\mathrm{1st}$ represents the loss function 
of the data and thus $M_2$, a bigger expressibility function, can lower this value; 
this is consistent with the general fact that a statistical model with more parameters 
tends to overfit to the observed data. 
The information criterion fixes this issue by introducing the penalty term; 
actually $\mathrm{QTIC}$ selects $M_1$ more frequently, meaning that $\mathrm{QTIC}$ 
can better capture the hidden function generating the data than 
$\mathrm{QTIC}_\mathrm{1st}$.

We can conduct model selection based on AIC as a reference. 
Because AIC is based on (classical) probability distributions, we measure the true and estimated quantum states and obtain the corresponding distributions. 
For model selection by AIC, we follow the same procedure depicted in Fig.~\ref{fig:num_proc} with some minor changes. 
More precisely, in Step 2, we use SLSQP to calculate the maximum likelihood estimator for the parameters, unlike the case of QTIC. 
In Steps 3 and 4, AIC is calculated and used for model selection instead of QTIC. 
Note that, while the measurement outcomes in Step 1 are used to construct the classical shadow for computing QTIC, the same measurement outcomes are used in Step 1 and represented as the empirical classical distribution in the case of AIC.
Table~\ref{tab:ranking_AIC} shows the number of selections of the model $M_1$ or $M_2$ out of 50 experiments based on the following three criteria. 
The first one, $\mathrm{CE}_\mathrm{true}$, is the negative classical cross entropy between the true distribution and the estimated distribution, where those classical distributions are obtained by the measurement of the $3$-fold tensor of the Pauli-6 POVM \cite{carrasquilla2019Reconstructing} for the true and the estimated states, respectively. 
The other two are the 1st term of AIC (denoted as $\mathrm{AIC}_\mathrm{1st}$) and AIC. 
The result indicates the similar tendency to the result of QTIC shown in Table~\ref{tab:ranking}. 
That is, $\mathrm{CE}_\mathrm{true}$ chooses $M_1$ more often than $M_2$; moreover, while $\mathrm{AIC}_\mathrm{1st}$ chooses $M_2$ more often, AIC corrects the bias of this criterion and thereby can choose the correct model $M_1$ with high probability $(47/50=94\%)$.

\begin{table}
  \parbox{.45\linewidth}{
  \centering
  \caption{The number of selecting $M_1$ or $M_2$ out of 50 experiments based on $\mathrm{QCE}_{\mathrm{true}}$, the $1^\mathrm{st}$ term of $\mathrm{QTIC}$ ($\mathrm{QTIC}_\mathrm{1st}$), and $\mathrm{QTIC}$.}
  \label{tab:ranking}
  \begin{tabular}{|l|c|c|c|}
      % \Hline %% ←
      \hline
      & $\mathrm{QCE}_{\mathrm{true}}$ & $\mathrm{QTIC}_\mathrm{1st}$ & $\mathrm{QTIC}$ \\
      \hline
      $M_1$ & 50 & 0 & 36\\
      \hline
      $M_2$ & 0 & 50 & 14\\
      \hline
      % \Hline %% ←
  \end{tabular}
  }
  \hfill
  \parbox{.45\linewidth}{
  \centering
  \caption{The number of selecting $M_1$ or $M_2$ out of 50 experiments based on $\mathrm{CE}_{\mathrm{true}}$, the $1^\mathrm{st}$ term of $\mathrm{AIC}$ ($\mathrm{AIC}_\mathrm{1st}$), and $\mathrm{AIC}$.}
  \label{tab:ranking_AIC}
  \begin{tabular}{|l|c|c|c|}
      % \Hline %% ←
      \hline
      & $\mathrm{CE}_{\mathrm{true}}$ & $\mathrm{AIC}_\mathrm{1st}$ & $\mathrm{AIC}$ \\
      \hline
      $M_1$ & 29 & 23 & 47\\
      \hline
      $M_2$ & 21 & 27 & 3\\
      \hline
      % \Hline %% ←
  \end{tabular}
  }
\end{table}

We finally remark that both $M_1$ and $M_2$ are singular models. 
Therefore, by definition, both QTIC and AIC are invalid to use, as discussed in Section \ref{sec:InformationCriteria}.
Actually, in computing QTIC, we observed that the calculated $\hat{J}^{\mathrm{emp}}_{Q}(\hat{\bm\theta}_Q)$ 
was degenerate, especially for $M_2$. 
However, we computed $\mathrm{QTIC}$ by using the pseudo-inverse of 
$\hat{J}^{\mathrm{emp}}_{Q}(\hat{\bm\theta}_Q)$, which is the technique often used 
in classical statistics. 
As seen in the above numerical simulation, 
%before, 
using the pseudo-inverse did not affect the performance badly for 
this small-scale experiment. 
Nevertheless, in general, comparing different information criteria (QTIC and AIC in our case) for singular models does not provide useful information as it is proven that such information criteria are not equal to the unbiased estimate for the generalization error anymore \cite{hagiwara2002Problem}; 
therefore, although from Tables~\ref{tab:ranking} and \ref{tab:ranking_AIC} it seems that AIC (which chooses $M_1$ with probability $47/50=94\%$) realizes better model selection than QTIC (which chooses $M_1$ with probability $36/50=72\%$), we cannot have such conclusion.

%%%%%%%%%%%%%%%%%%%%%%%%%%%%%%%%%%%%%%%%%%%%%%%%%%
%%%%%%%%%%%%%%%%%%%%%%%%%%%%%%%%%%%%%%%%%%%%%%%%%%
%%%%%%%%%%%%%%%%%%%%%%%%%%%%%%%%%%%%%%%%%%%%%%%%%%
\section{Conclusion}
\label{sec:conclusion}

In this work, we developed quantum information criteria for statistical model selection, 
in terms of the quantum relative entropy. 
This is quantum generalization of the classical information criteria, AIC and TIC, which 
evaluate an estimated model based on the KL divergence. 
Those quantum information criteria depend on the type of estimator of the quantum 
relative entropy; 
i.e., the log-likelihood ($\mathrm{QAIC}_{\mathrm{LL}}$) and the classical shadow 
($\mathrm{QTIC}_{\mathrm{shadow}}$ and $\mathrm{QAIC}_{\mathrm{shadow}}$). 
$\mathrm{QAIC}_{\mathrm{LL}}$ explicitly incorporates the optimality of the measurement 
into a model selection criterion, which does not appear in the classical information 
criteria. 
Also, $\mathrm{QTIC}_{\mathrm{shadow}}$, which is more suitable for practical use 
compared to $\mathrm{QAIC}_{\mathrm{LL}}$, was examined in the numerical experiment for a simple model selection 
problem.
Recall that the quantum information criteria derived in this paper are the measure for determining a quantum state, while there are other model selection techniques in quantum information technology. 
For instance, the quantum tomographic transfer function \cite{rehacek2015Determining} is the measure for determining a tomographically complete measurement.

As a concluding remark, we discuss some practical issues of the proposed quantum 
information criteria. 
The first issue is the computational complexity. 
For instance, $\mathrm{QTIC}_{\mathrm{shadow}}$ requires much computation of 
$\hat{I}^{\mathrm{emp}}_{Q}(\hat{\bm\theta}_Q)$ and 
$\hat{J}^{\mathrm{emp}}_{Q}(\hat{\bm\theta}_Q)$. 
As pointed out in Section \ref{sec:numDemo}, the derivatives of these matrices 
have to be computed classically using the $n$ classical snapshots, instead of using 
the parameter shift rule on quantum computers. 
However, the complexity of this computation for the derivatives scales exponentially 
with the system size, and thus $\mathrm{QTIC}_{\mathrm{shadow}}$ faces the practical 
limit as it is. 
The second is on the problem of handling singular models. 
The quantum information criteria derived in this paper cannot be applied to singular 
models, because the Cram\'er-Rao framework does not hold in this case, as mentioned in 
Section \ref{sec:InformationCriteria}. 
Thus, they are likely to be invalid to use for parametric quantum states with many 
parameters. 
However, there should be a chance to design a quantum information criterion that can 
work even for singular models; actually, there already exists a Bayesian-based classical 
information criterion called WAIC \cite{watanabe2010Asymptotic} which is applicable 
for singular models. 
The last issue is on the measurement procedures. 
We assume that the measurement data are obtained by tomographically complete measurements, 
like random Pauli measurements, for simplifying the derivation of the quantum information 
criteria. 
However, it is not optimal for the purpose of parameter estimation. 
In fact, depending on the models to be compared, the measurement may not need to be 
tomographically complete, as noted at the beginning of Section \ref{subsec:qic_ll}. 
Furthermore, there is often the case where a tomographically complete measurement is 
impossible. 
It is worth investigating if we could compute the quantum information criteria without 
tomographically complete measurements. 
By solving the above practical issues, we expect that the proposed quantum information 
criteria can function as a relevant and practical model selection method for quantum 
state estimation problems.

%%%%%%%%%%%%%%%%%%%%%%%%%%%%%%%%%%%%%%%%%%%%%%%%%%
%%%%%%%%%%%%%%%%%%%%%%%%%%%%%%%%%%%%%%%%%%%%%%%%%%
\section*{Acknowledgments}
This work was supported by MEXT Quantum Leap Flagship Program (MEXT Q-LEAP) Grants No. JPMXS0118067285 and No. JPMXS0120319794 and JST SPRING Grant No. JPMJSP2123.

%%%%%%%%%%%%%%%%%%%%%%%%%%%%%%%%%%%%%%%%%%%%%%%%%%
%%%%%%%%%%%%%%%%%%%%%%%%%%%%%%%%%%%%%%%%%%%%%%%%%%
\section*{Data availability}
The data that support the findings of this study are available upon request from the authors.

\appendix
%%%%%%%%%%%%%%%%%%%%%%%%%%%%%%%%%%%%%%%%%%%%%%%%%%
%%%%%%%%%%%%%%%%%%%%%%%%%%%%%%%%%%%%%%%%%%%%%%%%%%
\section{Details of the bias derivation}
\label{app:derivation}
We give the detailed calculations of each term of the bias, $D_1$, $D_2$, and $D_3$, for the log-likelihood approach in \ref{app:derivation_LL}, and for the classical shadow approach in \ref{app:derivation_shadow}. 
The symbol $\approx$, such as in Eq.~(\ref{eq:LL_D3_eta_Taylor}), represents the approximation to the second order and ignores the higher order. 

\subsection{Log-likelihood approach}\label{app:derivation_LL}
\noindent\textbf{Calculation of $D_2$:} First we find 
\begin{eqnarray}
  D_2 &= \mathbb{E}_{g(\bm x_n)} \left[ \ell_{\mathrm{LL}}(\bm X_n ; \bm\theta_0) - n \Tr \left( \rho \log \sigma(\bm\theta_0) \right) \right] \nonumber\\
  &= n \left( \mathbb{E}_{g(\bm x_n)} \left[ \frac{1}{n} \sum_{\alpha=1}^n \log \Tr \left( \Pi_{X_\alpha} \sigma(\bm\theta_0) \right) \right] - \Tr \left( \rho \log \sigma(\bm\theta_0) \right) \right) \nonumber\\
  &= n \left( \mathbb{E}_{g(\bm x_n)} \left[ \sum_m \sum_{\alpha=1}^n \frac{\delta(m-X_\alpha)}{n} \log \Tr \left( \Pi_m \sigma(\bm\theta_0) \right) \right] - \Tr \left( \rho \log \sigma(\bm\theta_0) \right) \right) \nonumber\\
  &= n \left( \sum_m \Tr \left( \Pi_{m} \rho \right) \log \Tr \left( \Pi_{m} \sigma(\bm\theta_0) \right) - \Tr \left( \rho \log \sigma(\bm\theta_0) \right) \right) \nonumber\\
  &= n \left( H(g(z)\|\Tr \left( \Pi_{z} \sigma(\bm\theta_0) \right)) - \Tr \left( \rho \log \sigma(\bm\theta_0) \right) \right). \label{eq:LL_D2}
\end{eqnarray}
The fourth equality uses the fact that $\mathbb{E}_{g(\bm x_n)} \left[ \sum_{\alpha=1}^n \frac{\delta(m-X_\alpha)}{n} \right] = \Tr \left( \Pi_{m} \rho \right), \forall m$.
In the last equality, we use $g(z) = \Tr \left( \Pi_{z} \rho \right)$ and the negative 
classical cross entropy $H$ defined in Eq.~(\ref{eq:negativeCCE}). 
\\

\noindent\textbf{Calculation of $D_3$:} 
Define $\eta(\bm\theta) = \Tr \left( \rho \log \sigma(\bm\theta) \right)$. 
Then we have 
\begin{eqnarray}\label{eq:LL_D3_eta}
  D_3 &= n \mathbb{E}_{g(\bm x_n)} \left[ \Tr \left( \rho \log \sigma(\bm\theta_0) \right) - \Tr \left( \rho \log \sigma(\hat{\bm\theta}_C) \right) \right] \nonumber\\
  &= n \mathbb{E}_{g(\bm x_n)} \left[ \eta(\bm\theta_0) - \eta(\hat{\bm\theta}_C) \right].
\end{eqnarray}
Here, the Taylor expansion of $\eta(\hat{\bm\theta}_C)$ around $\bm\theta_0$ gives
\begin{eqnarray}\label{eq:LL_D3_eta_Taylor}
    \eta(\hat{\bm\theta}_C) &\approx \eta(\bm\theta_0) + (\hat{\bm\theta}_C - \bm\theta_0)^T \frac{\partial \eta(\bm\theta_0)}{\partial \bm\theta} + \frac{1}{2} (\hat{\bm\theta}_C - \bm\theta_0)^T \frac{\partial^2 \eta(\bm\theta_0)}{\partial \bm\theta \partial \bm\theta^T} (\hat{\bm\theta}_C - \bm\theta_0) \nonumber\\
    &= \eta(\bm\theta_0) - \frac{1}{2} (\hat{\bm\theta}_C - \bm\theta_0)^T J_Q(\bm\theta_0) (\hat{\bm\theta}_C - \bm\theta_0),
\end{eqnarray}
where for $i,j = 1, ..., p$,
\begin{equation}\label{eq:J_Q}
    J_{Q;ij}(\bm\theta_0) = - \frac{\partial^2 \eta(\bm\theta_0)}{\partial \theta_i \partial \theta_j}
    = - \Tr \left( \rho \left. \frac{\partial^2 \log \sigma(\bm\theta)}{\partial \theta_i \partial \theta_j} \right|_{\bm\theta = \bm\theta_0} \right).
\end{equation}
Hence, according to Eqs.~(\ref{eq:LL_D3_eta}) and (\ref{eq:LL_D3_eta_Taylor}),
\begin{eqnarray}
  D_3 &\approx \frac{n}{2} \mathbb{E}_{g(\bm x_n)} \left[ (\hat{\bm\theta}_C - \bm\theta_0)^T J_Q(\bm\theta_0) (\hat{\bm\theta}_C - \bm\theta_0) \right] \nonumber\\
  &= \frac{n}{2} \mathbb{E}_{g(\bm x_n)} \left[ \Tr \left( J_Q(\bm\theta_0) (\hat{\bm\theta}_C - \bm\theta_0) (\hat{\bm\theta}_C - \bm\theta_0)^T  \right) \right] \nonumber\\
  &= \frac{n}{2} \Tr \left( J_Q(\bm\theta_0) \mathbb{E}_{g(\bm x_n)} \left[ (\hat{\bm\theta}_C - \bm\theta_0) (\hat{\bm\theta}_C - \bm\theta_0)^T \right] \right) \nonumber\\
  &\approx \frac{1}{2} \Tr \left( J_Q(\bm\theta_0) J_C(\bm\theta_0)^{-1} I_C(\bm\theta_0) J_C(\bm\theta_0)^{-1} \right), 
\label{eq:LL_D3}
\end{eqnarray}
where
\begin{eqnarray}
  I_{C;ij}(\bm\theta_0) &= \mathbb{E}_{g(x)} \left[ \left. \frac{\partial \log \Tr \left( \Pi_X \sigma(\bm\theta) \right) }{\partial \theta_i} \frac{\partial \log \Tr \left( \Pi_X \sigma(\bm\theta) \right)}{\partial \theta_j} \right|_{\bm\theta = \bm\theta_0} \right], \label{eq:LL_I_C}
  \\
  J_{C;ij}(\bm\theta_0) &= - \mathbb{E}_{g(x)} \left[ \left. \frac{\partial^2 \log \Tr \left( \Pi_X \sigma(\bm\theta) \right)}{\partial \theta_i \partial \theta_j} \right|_{\bm\theta = \bm\theta_0} \right]. \label{eq:LL_J_C}
\end{eqnarray}
The last approximation in Eq.~(\ref{eq:LL_D3}) is obtained by the asymptotic 
normality of the maximum likelihood estimator $\hat{\bm\theta}_C$; 
see~\ref{subapp:asymptotic_normality_classical}.
\\

\noindent\textbf{Calculation of $D_1$:} Here we focus on
\begin{equation}
  D_1 = \mathbb{E}_{g(\bm x_n)} \left[ \ell_{\mathrm{LL}}(\bm X_n ; \hat{\bm\theta}_C) - \ell_{\mathrm{LL}}(\bm X_n ; \bm\theta_0) \right].
\end{equation}
The Taylor expansion of $\ell_{\mathrm{LL}}(\bm X_n ;\bm\theta)$ around $\hat{\bm\theta}_C$ gives
\begin{eqnarray}
  \fl \ell_{\mathrm{LL}}(\bm X_n ;\bm\theta) &\approx \ell_{\mathrm{LL}}(\bm X_n ;\hat{\bm\theta}_C) + (\bm\theta - \hat{\bm\theta}_C)^T \frac{\partial \ell_{\mathrm{LL}}(\bm X_n ;\hat{\bm\theta}_C)}{\partial \bm\theta} + \frac{1}{2} (\bm\theta - \hat{\bm\theta}_C)^T \frac{\partial^2 \ell_{\mathrm{LL}}(\bm X_n ;\hat{\bm\theta}_C)}{\partial \bm\theta \partial \bm\theta^T} (\bm\theta - \hat{\bm\theta}_C) \nonumber\\
  \fl &= \ell_{\mathrm{LL}}(\bm X_n ;\hat{\bm\theta}_C) + \frac{1}{2} (\bm\theta - \hat{\bm\theta}_C)^T \frac{\partial^2 \ell_{\mathrm{LL}}(\bm X_n ;\hat{\bm\theta}_C)}{\partial \bm\theta \partial \bm\theta^T} (\bm\theta - \hat{\bm\theta}_C) \nonumber\\
  \fl &\rightarrow \ell_{\mathrm{LL}}(\bm X_n ;\hat{\bm\theta}_C) - \frac{n}{2} (\bm\theta - \hat{\bm\theta}_C)^T J_C(\bm\theta_0) (\bm\theta - \hat{\bm\theta}_C).
\end{eqnarray}
At the last line, we used the fact that the quantity
\begin{equation*}
  - \frac{1}{n} \frac{\partial^2 \ell_{\mathrm{LL}}(\bm X_n ;\hat{\bm\theta}_C)}{\partial \bm\theta \partial \bm\theta^T} 
  = - \frac{1}{n} \frac{\partial^2 \left(\sum_{\alpha=1}^n \log \Tr \left( \Pi_{x_\alpha} \sigma(\hat{\bm\theta}_C) \right)\right)}{\partial \bm\theta \partial \bm\theta^T}
\end{equation*}
converges in probability to $J_C(\bm\theta_0)$ defined in Eq.~(\ref{eq:LL_J_C}) 
when $n \rightarrow \infty$. 
Using this result, we obtain
\begin{eqnarray}
  D_1 &\approx \frac{n}{2} \mathbb{E}_{g(\bm{x}_n)} \left[ (\bm\theta_0 - \hat{\bm\theta}_C)^T J_C(\bm\theta_0) (\bm\theta_0 - \hat{\bm\theta}_C) \right] \nonumber\\
  &= \frac{n}{2} \mathbb{E}_{g(\bm{x}_n)} \left[ \Tr \left( J_C(\bm\theta_0) (\bm\theta_0 - \hat{\bm\theta}_C) (\bm\theta_0 - \hat{\bm\theta}_C)^T \right) \right] \nonumber\\
  &= \frac{n}{2} \Tr \left( J_C(\bm\theta_0) \mathbb{E}_{g(\bm{x}_n)} \left[ (\hat{\bm\theta}_C - \bm\theta_0) (\hat{\bm\theta}_C - \bm\theta_0)^T \right] \right) \nonumber\\
  &\approx \frac{1}{2} \Tr \left( I_C(\bm\theta_0) J_C(\bm\theta_0)^{-1} \right). \label{eq:LL_D1}
\end{eqnarray}
At the last approximation we utilize the asymptotic normality of the maximum 
likelihood estimator $\hat{\bm\theta}_C$, provided in 
\ref{subapp:asymptotic_normality_classical}.

\subsection{Classical shadow approach}\label{app:derivation_shadow}
\noindent\textbf{Calculation of $D_2$:} 
First, we find 
\begin{eqnarray}
  D_2 &= \mathbb{E}_{g(\hat{\bm\rho}_n)} \left[ \ell_{\mathrm{shadow}}(\hat{\bm\rho}_n; \hat{\bm\theta}_0) - n \Tr \left( \rho \log \sigma(\bm\theta_0) \right) \right] \nonumber\\
  &= n \left( \mathbb{E}_{g(\hat{\bm\rho}_n)} \left[ \Tr \left( \left( \frac{1}{n} \sum_{i=1}^n \hat{\rho}_i \right) \log \sigma(\bm\theta_0) \right) \right] - \Tr \left( \rho \log \sigma(\bm\theta_0) \right) \right) \nonumber\\
  &= n \left( \Tr \left( \rho \log \sigma(\bm\theta_0) \right) - \Tr \left( \rho \log \sigma(\bm\theta_0) \right) \right) = 0. 
\label{eq:shadow_D2}
\end{eqnarray}
\mbox{}
\\

\noindent\textbf{Calculation of $D_3$:} 
Define $\eta(\bm\theta) = \Tr \left( \rho \log \sigma(\bm\theta) \right)$. 
Then $D_3$ is expressed as 
\begin{eqnarray}\label{eq:shadow_D3_eta}
  D_3 &= n \mathbb{E}_{g(\hat{\bm\rho}_n)} \left[ \Tr \left( \rho \log \sigma(\bm\theta_0) \right) - \Tr \left( \rho \log \sigma(\hat{\bm\theta}_Q) \right) \right] \nonumber\\
  &= n \mathbb{E}_{g(\hat{\bm\rho}_n)} \left[ \eta(\bm\theta_0) - \eta(\hat{\bm\theta}_Q) \right].
\end{eqnarray}
Here, the Taylor expansion of $\eta(\hat{\bm\theta}_Q)$ around $\bm\theta_0$ gives
\begin{eqnarray}\label{eq:shadow_D3_eta_Taylor}
    \eta(\hat{\bm\theta}_Q) &\approx \eta(\bm\theta_0) + (\hat{\bm\theta}_Q - \bm\theta_0)^T \frac{\partial \eta(\bm\theta_0)}{\partial \bm\theta} + \frac{1}{2} (\hat{\bm\theta}_Q - \bm\theta_0)^T \frac{\partial^2 \eta(\bm\theta_0)}{\partial \bm\theta \partial \bm\theta^T} (\hat{\bm\theta}_Q - \bm\theta_0) \nonumber\\
    &= \eta(\bm\theta_0) - \frac{1}{2} (\hat{\bm\theta}_Q - \bm\theta_0)^T J_Q(\bm\theta_0) (\hat{\bm\theta}_Q - \bm\theta_0),
\end{eqnarray}
where $J_Q(\bm\theta_0)$ is equivalent to Eq.~(\ref{eq:J_Q}) obtained in the case of 
the estimator using the log-likelihood. 
From Eqs.~(\ref{eq:shadow_D3_eta}) and (\ref{eq:shadow_D3_eta_Taylor}), we have 
\begin{eqnarray*}
  D_3 &\approx \frac{n}{2} \mathbb{E}_{g(\hat{\bm\rho}_n)} \left[ (\hat{\bm\theta}_Q - \bm\theta_0)^T J_Q(\bm\theta_0) (\hat{\bm\theta}_Q - \bm\theta_0) \right] \\
  &= \frac{n}{2} \mathbb{E}_{g(\hat{\bm\rho}_n)} \left[ \Tr \left( J(\bm\theta_0) (\hat{\bm\theta}_Q - \bm\theta_0) (\hat{\bm\theta}_Q - \bm\theta_0)^T \right) \right] \\
    &= \frac{n}{2} \Tr \left( J_Q(\bm\theta_0) \mathbb{E}_{g(\hat{\bm\rho}_n)} \left[ (\hat{\bm\theta}_Q - \bm\theta_0) (\hat{\bm\theta}_Q - \bm\theta_0)^T \right] \right).
\end{eqnarray*}
The covariance matrix of $\hat{\bm\theta}_Q$ around $\bm\theta_0$ appears in the 
last line; unlike the case of likelihood approach where Eq.~(\ref{eq:LL_D3}) holds, 
the estimator $\hat{\bm\theta}_Q$ does not satisfy the standard asymptotic normality 
for classical maximum likelihood estimators. 
Instead, using the asymptotic normality for the classical shadow proven in 
\ref{subapp:asymptotic_normality_shadow}, we obtain
\begin{eqnarray}
  D_3 &\approx \frac{1}{2} \Tr \left( J_Q(\bm\theta_0) J_Q(\bm\theta_0)^{-1} I_Q(\bm\theta_0) J_Q(\bm\theta_0)^{-1} \right) \nonumber\\
  &= \frac{1}{2} \Tr \left( I_Q(\bm\theta_0) J_Q(\bm\theta_0)^{-1} \right), \label{eq:shadow_D3}
\end{eqnarray}
where
\begin{equation}\label{eq:I_Q_shadow}
  I_{Q;ij}(\bm\theta_0) = \mathbb{E}_{g(\hat{\rho})} \left[ \Tr \left( \hat{\rho} \left. \frac{\partial \log \sigma(\bm\theta)}{\partial \theta_i} \right|_{\bm\theta = \bm\theta_0}  \right) \Tr \left( \hat{\rho} \left. \frac{\partial \log \sigma(\bm\theta)}{\partial \theta_j} \right|_{\bm\theta = \bm\theta_0}  \right) \right].
\end{equation}
\mbox{}
\\

\noindent\textbf{Calculation of $D_1$:} 
Next we focus on 
\begin{equation}
  D_1 = \mathbb{E}_{g(\hat{\bm\rho}_n)} \left[ \ell_{\mathrm{shadow}}(\hat{\bm\rho}_n; \hat{\bm\theta}_Q) - \ell_{\mathrm{shadow}}(\hat{\bm\rho}_n; \hat{\bm\theta}_0) \right].
\end{equation}
Again, the Taylor expansion of $\ell_{\mathrm{shadow}}(\hat{\bm\rho}_n ; \bm\theta)$ 
around $\hat{\bm\theta}_Q$ gives
\begin{eqnarray}
  \ell_{\mathrm{shadow}}(\hat{\bm\rho}_n ; \bm\theta) &\approx \ell_{\mathrm{shadow}}(\hat{\bm\rho}_n ; \hat{\bm\theta}_Q) + (\bm\theta - \hat{\bm\theta}_Q)^T \frac{\partial \ell_{\mathrm{shadow}}(\hat{\bm\rho}_n ; \hat{\bm\theta}_Q)}{\partial \bm\theta} \nonumber\\
  & \quad \, + \frac{1}{2} (\bm\theta - \hat{\bm\theta}_Q)^T \frac{\partial^2 \ell_{\mathrm{shadow}}(\hat{\bm\rho}_n ; \hat{\bm\theta}_Q)}{\partial \bm\theta \partial \bm\theta^T} (\bm\theta - \hat{\bm\theta}_Q) \nonumber\\
  &= \ell_{\mathrm{shadow}}(\hat{\bm\rho}_n ; \hat{\bm\theta}_Q) + \frac{1}{2} (\bm\theta - \hat{\bm\theta}_Q)^T \frac{\partial^2 \ell_{\mathrm{shadow}}(\hat{\bm\rho}_n ; \hat{\bm\theta}_Q)}{\partial \bm\theta \partial \bm\theta^T} (\bm\theta - \hat{\bm\theta}_Q) \nonumber\\
  &\rightarrow \ell_{\mathrm{shadow}}(\hat{\bm\rho}_n ; \hat{\bm\theta}_Q) - \frac{n}{2} (\bm\theta - \hat{\bm\theta}_Q)^T J_Q(\bm\theta_0) (\bm\theta - \hat{\bm\theta}_Q).
\end{eqnarray}
The convergence in the last line is guaranteed as follows. 
That is, 
\begin{equation*}
  \frac{1}{n} \frac{\partial^2 \ell_{\mathrm{shadow}}(\hat{\bm\rho}_n ; \hat{\bm\theta}_Q)}{\partial \bm\theta \partial \bm\theta^T} = \frac{1}{n} \frac{\partial^2 \left( n \Tr \left( \left( \frac{1}{n} \sum_{\alpha=1}^n \hat{\rho}_\alpha \right) \log \sigma(\hat{\bm\theta}_Q) \right) \right)}{\partial \bm\theta \partial \bm\theta^T}
\end{equation*}
converges in probability to $J_Q(\bm\theta_0)$ when $n \rightarrow \infty$, based on 
the fact that $\frac{1}{n} \sum_{\alpha=1}^n \hat{\rho}_\alpha$ and $\hat{\bm\theta}_Q$  
converges to $\rho$ and $\bm\theta_0$, respectively. 
From this approximation, we have
\begin{eqnarray}
  D_1 &\approx \frac{n}{2} \mathbb{E}_{g(\hat{\bm\rho}_n)} \left[ (\bm\theta_0 - \hat{\bm\theta}_Q)^T J_Q(\bm\theta_0) (\bm\theta_0 - \hat{\bm\theta}_Q) \right] \nonumber\\
  &= \frac{n}{2} \mathbb{E}_{g(\hat{\bm\rho}_n)} \left[ \Tr \left( J_Q(\bm\theta_0) (\bm\theta_0 - \hat{\bm\theta}_Q) (\bm\theta_0 - \hat{\bm\theta}_Q)^T \right) \right] \nonumber\\
  &= \frac{n}{2} \Tr \left( J_Q(\bm\theta_0) \mathbb{E}_{g(\hat{\bm\rho}_n)} \left[ (\hat{\bm\theta}_Q - \bm\theta_0) (\hat{\bm\theta}_Q - \bm\theta_0)^T \right] \right) \nonumber\\
  &\approx \frac{1}{2} \Tr \left( I_Q(\bm\theta_0) J_Q(\bm\theta_0)^{-1} \right) \label{eq:shadow_D1}
\end{eqnarray}
In the last approximation we again utilized the asymptotic normality of the estimator 
$\hat{\bm\theta}_Q$ proven in \ref{subapp:asymptotic_normality_shadow}.

%%%%%%%%%%%%%%%%%%%%%%%%%%%%%%%%%%%%%%%%%%%%%%%%%%
\section{Asymptotic normality}
\label{app:asymptotic_normality}

In this appendix, the asymptotic properties of the estimators $\hat{\bm\theta}_C$ and 
$\hat{\bm\theta}_Q$ are investigated, which we utilize for the derivation of quantum 
information criteria in Section~\ref{sec:quantumInformationCriteria}.

\subsection{Asymptotic normality for random variables}
\label{subapp:asymptotic_normality_classical}

The data $\bm x_n = \{ x_1, x_2, ..., x_n \}$ are observed according to the 
distribution $g(x)$. 
Assume that the regularity condition holds for the density function $f(x|\bm\theta)$ 
and $\bm\theta_0$ is a unique solution of
\begin{equation*}
  \int g(x) \frac{\partial \log f(x|\bm\theta)}{\partial \bm\theta} dx = \bm 0.
\end{equation*}
Then, the following statements hold with respect to the maximum likelihood estimator 
$\hat{\bm\theta}_C$:
\begin{enumerate}
  \item The maximum likelihood estimator $\hat{\bm\theta}_C$ converges in probability to $\bm\theta_0$ as $n \rightarrow \infty$.
  \item When $n \rightarrow \infty$,
  \begin{equation*}
      \sqrt{n} (\hat{\bm\theta}_C - \bm\theta_0) \rightarrow_d \mathcal{N}_p \left(\bm 0, J_C(\bm\theta_0)^{-1} I_C(\bm\theta_0) J_C(\bm\theta_0)^{-1} \right),
  \end{equation*}
  where $\mathcal{N}_p$ represents the Gaussian distribution for $p$-dimensional random variables, and 
  $I(\bm\theta)$ and $J(\bm\theta)$ are $p \times p$ matrices given by
  \begin{eqnarray*}
      I_{C;ij}(\bm\theta) &= \mathbb{E}_{g(z)} \left[ \frac{\partial \log f(Z|\bm\theta)}{\partial \theta_i} \frac{\partial \log f(Z|\bm\theta)}{\partial \theta_j} \right], \\
      J_{C;ij}(\bm\theta) &= - \mathbb{E}_{g(z)} \left[ \frac{\partial^2 \log f(Z|\bm\theta)}{\partial \theta_i \partial \theta_j} \right].
  \end{eqnarray*}
  The convergence in distribution is denoted by $\rightarrow_d$.
\end{enumerate}
%\begin{proof}
%  See Section~3.3.5 in Ref.~\cite{konishi2008information}.
%\end{proof}

\subsection{Asymptotic normality for the classical shadow}
\label{subapp:asymptotic_normality_shadow}

The data $\hat{\bm\rho}_n = \{ \hat{\rho}_1, \hat{\rho}_2, ..., \hat{\rho}_n \}$ is the 
classical shadow of $\rho$, i.e., $\mathbb{E}_{g(\hat{\rho})}[\hat{\rho}] = \rho$ with 
probability $g(\hat{\rho}) = \bra{\hat{b}} U \rho U^\dagger \ket{\hat{b}}$, as described 
in Section~\ref{subsec:qic_shadow}. 
Assume that the regularity condition holds for the function $\Tr \left( \hat{\rho} \log \sigma(\bm\theta) \right)$ and
$\bm\theta_0$ is a unique solution of
\begin{equation*}
    \Tr \left( \rho \frac{\partial \log \sigma(\bm\theta)}{\partial \bm\theta} \right) = \bm 0.
\end{equation*}
Then, the estimator
\begin{equation*}
  \hat{\bm\theta}_Q = \argmax_{\bm\theta \in \Theta} \sum_{\alpha=1}^n \Tr \left( \hat{\rho}_\alpha \log \sigma(\bm\theta) \right)
\end{equation*}
satisfies the following properties: 
\begin{enumerate}
    \item The estimator $\hat{\bm\theta}_Q$ converges in probability to $\bm\theta_0$ as 
    $n \rightarrow \infty$.
    \item When $n \rightarrow \infty$,
    \begin{equation*}
        \sqrt{n} (\hat{\bm\theta}_Q - \bm\theta_0) \rightarrow_d \mathcal{N}_p\left(\bm 0, J_Q(\bm\theta_0)^{-1} I_Q(\bm\theta_0) J_Q(\bm\theta_0)^{-1} \right),
    \end{equation*}
    where $I_Q(\bm\theta)$ and $J_Q(\bm\theta)$ are $p \times p$ matrices given by:
    \begin{eqnarray*}
        I_{Q;ij}(\bm\theta) &= \mathbb{E}_{g(\hat{\rho})} \left[ \Tr \left( \hat{\rho} \frac{\partial \log \sigma(\bm\theta)}{\partial \theta_i} \right) \Tr \left( \hat{\rho} \frac{\partial \log \sigma(\bm\theta)}{\partial \theta_j} \right) \right], \\
        J_{Q;ij}(\bm\theta) &= - \Tr \left( \rho \frac{\partial^2 \log \sigma(\bm\theta)}{\partial \theta_i \partial \theta_j} \right).
    \end{eqnarray*}
\end{enumerate}
\begin{proof}
\begin{enumerate}
  \item
  Let us define
  \begin{equation*}
        \ell_{\mathrm{shadow}}(\hat{\bm\rho}_n ; \bm\theta) = \sum_{\alpha=1}^n \Tr \left( \hat{\rho}_\alpha \log \sigma(\bm\theta) \right).
  \end{equation*}
  Since $\ell_{\mathrm{shadow}}(\hat{\bm\rho}_n ; \bm\theta)/n$ is the sample mean of $\Tr \left( \hat{\rho} \log \sigma(\bm\theta) \right)$, by the law of large numbers, it holds that
  \begin{equation}
        \frac{1}{n} \ell_{\mathrm{shadow}}(\hat{\bm\rho}_n ; \bm\theta) \rightarrow_p S(\rho \| \sigma(\bm\theta)),
  \end{equation}
  where $S(\cdot \| \cdot)$ is the negative QCE defined in Eq.~(\ref{eq:negativeQCE}) and $\rightarrow_p$ means the convergence in probability. 
  Therefore, $\hat{\bm\theta}_Q$ converges in probability to 
  $\bm\theta_0 = \argmax_{\bm\theta}S(\rho \| \sigma(\bm\theta))$.
  \item 
  The Taylor expansion of the first derivative of $\ell_{\mathrm{shadow}}(\hat{\bm\rho}_n ; \hat{\bm\theta}_Q)$ around $\bm\theta_0$ gives
  \begin{equation}
      \fl 0 \approx \frac{\partial}{\partial \theta_i} \ell_{\mathrm{shadow}}(\hat{\bm\rho}_n ; \bm\theta_0) 
         + \sum_{j=1}^p (\hat{\theta}_{Q;j} - \theta_{0;j}) \frac{\partial^2}{\partial \theta_i \partial \theta_j} \ell_{\mathrm{shadow}}(\hat{\bm\rho}_n ; \bm\theta_0), \quad i = 1, ..., p,
  \label{eq:asympNorm_Taylor}
  \end{equation}
  where we write $\hat{\theta}_{Q;j}$ and $\theta_{0;j}$ for the $j$-th element of $\hat{\bm\theta}_{Q}$ and $\bm\theta_{0}$, respectively.
  Let $\ell'_{\mathrm{shadow}}(\bm\theta_0)$ stand for the vector whose $i$-th element is $\partial_i \ell_{\mathrm{shadow}}(\hat{\bm\rho}_n ; \bm\theta_0)$.
  Moreover, we will denote by $n \hat{J}^{\mathrm{emp}}_Q(\bm\theta_0)$ the matrix whose $(i,j)$ element is $- \partial_i \partial_j \ell_{\mathrm{shadow}}(\hat{\bm\rho}_n ; \bm\theta_0)$.
  Then, Eq.~(\ref{eq:asympNorm_Taylor}) becomes
  \begin{equation}\label{eq:asympNorm_Taylor_abbrv}
      \sqrt{n} (\hat{\bm\theta}_Q - \bm\theta_0) \approx \hat{J}^{\mathrm{emp}}_Q(\bm\theta_0)^{-1} \frac{1}{\sqrt{n}} \ell'_{\mathrm{shadow}}(\bm\theta_0).
  \end{equation}
  By the law of large numbers, when $n \rightarrow \infty$, it can be shown that
  \begin{eqnarray}
        \hat{J}^{\mathrm{emp}}_Q(\bm\theta_0) &= - \frac{1}{n} \frac{\partial^2 \ell_{\mathrm{shadow}}(\hat{\bm\rho}_n ; \bm\theta_0)}{\partial \bm\theta \partial \bm\theta^T} \nonumber\\
        &= - \frac{1}{n} \sum_{\alpha=1}^n \left. \frac{\partial^2}{\partial \bm\theta \partial \bm\theta^T} \Tr \left( \hat{\rho}_\alpha \log \sigma(\bm\theta) \right) \right|_{\bm\theta = \bm\theta_0}
        \rightarrow_p J_Q(\bm\theta_0),
        \label{eq:asympNorm_J_Q}
  \end{eqnarray}
  Moreover, according to the multivariate central limit theorem for a $p$-dimensional random vector 
  $X_\alpha = \left. \frac{\partial}{\partial \bm\theta} \Tr \left( \hat{\rho}_\alpha \log \sigma(\bm\theta) \right) \right|_{\bm\theta = \bm\theta_0}$, we have
  \begin{eqnarray}
      \frac{1}{\sqrt{n}} \ell'_{\mathrm{shadow}}(\bm\theta_0) &= \frac{1}{\sqrt{n}} \frac{\partial \ell_{\mathrm{shadow}}(\bm\theta_0)}{\partial \bm\theta} \nonumber\\
      &= \frac{1}{\sqrt{n}} \sum_{\alpha=1}^n \left. \frac{\partial}{\partial \bm\theta} \Tr \left( \hat{\rho}_\alpha \log \sigma(\bm\theta) \right) \right|_{\bm\theta = \bm\theta_0}
      \rightarrow_d \mathcal{N}_p(\bm 0, I_Q(\bm\theta_0)).
      \label{eq:asympNorm_I_Q}
  \end{eqnarray}
  Therefore, from Eqs.~(\ref{eq:asympNorm_Taylor_abbrv}), (\ref{eq:asympNorm_J_Q}), and (\ref{eq:asympNorm_I_Q}), we use the Slutsky's theorem to obtain
  \begin{equation*}
    \sqrt{n} (\hat{\bm\theta}_Q - \bm\theta_0) \rightarrow_d \mathcal{N}_p(\bm 0, J_Q(\bm\theta_0)^{-1} I_Q(\bm\theta_0) J_Q(\bm\theta_0)^{-1}),
  \end{equation*}
  when $n \rightarrow \infty$.
\end{enumerate}
\end{proof}

%%%%%%%%%%%%%%%%%%%%%%%%%%%%%%%%%%%%%%%%%%%%%%%%%%
\section{Proof of convergence of the maximum likelihood estimator}
\label{app:proof_of_convergence}

The motivation for the use of log-likelihood in Section~\ref{subsec:qic_ll} is based 
on the fact that the maximum likelihood estimator $\hat{\bm\theta}_C$ given in 
Eq.~(\ref{eq:theta_C}) converges to $\bm\theta_0$, which is a solution to 
$\Tr \left( \rho \partial_{\bm\theta} \log \sigma(\bm\theta) \right) = \bm 0$ and 
thus can be a maximizer of the negative QCE (\ref{eq:negativeQCE}). 
Here we give a proof for this argument.

The proof consists of two steps. 
First, we show that $\bm\theta_0$ maximizes the expected log-likelihood or the negative 
classical cross entropy (\ref{eq:negativeCCE}) as well as the negative QCE; second, we 
use the consistency of $\hat{\bm\theta}_C$.
We also refer to Ref.~\cite{shangnan2021Quantum} for the first step. 
Below, we assume the regularity condition on $\sigma(\bm\theta)$ and the realizable 
case, i.e., there exists $\bm\theta_0$ such that $\rho = \sigma(\bm\theta_0)$.

The maximum of the negative QCE $S(\rho\|\sigma(\bm\theta))$ can be achieved only when 
$\sigma(\bm\theta) = \rho$ due to the non-negativity of the quantum relative entropy 
(\ref{eq:qre}); that is, $D(\rho_1\|\rho_2)$ is zero if and only if $\rho_1 = \rho_2$. 
Thus,
\begin{equation}\label{eq:app_theta_0_S}
  \bm\theta_0 = \argmax_{\bm\theta \in \Theta} S(\rho\|\sigma(\bm\theta)).
\end{equation}
We next consider the log-likelihood function $\ell_{\mathrm{LL}}$ defined in 
Eq.~(\ref{eq:theta_C}). 
In the limit $n \rightarrow \infty$, we have 
\begin{eqnarray}
  \frac{1}{n} \ell_{\mathrm{LL}}(\bm{x}_n; \bm\theta) &\rightarrow_p \sum_m \Tr \left( \Pi_{m} \rho \right) \log \Tr \left( \Pi_{m} \sigma(\bm\theta) \right) \\
  &= H \left( \Tr \left( \Pi_{Z} \rho \right) \| \Tr \left( \Pi_{Z} \sigma(\bm\theta) \right) \right),
\end{eqnarray}
by the law of large numbers, where $H(\cdot\|\cdot)$ is the negative classical cross entropy 
(\ref{eq:negativeCCE}) and $\rightarrow_p$ means the convergence in probability. 
Similarly, the maximum of $H$ can be achieved only when 
\begin{equation}\label{eq:app_eq_prob}
  \Tr \left( \Pi_{m} \rho \right) = \Tr \left( \Pi_{m} \sigma(\bm\theta) \right) \quad \mbox{for all } m,
\end{equation}
due to the non-negativity of the KL divergence (\ref{eq:KL}); that is, $KL(p_1\|p_2)$ 
is zero if and only if $p_1 = p_2$. 
Because $\{\Pi_m\}$ is the tomographically complete measurement, the above equation implies that $\sigma(\bm\theta) = \rho$ for the state which maximizes the negative classical cross entropy $H$.
Since $\rho = \sigma(\bm\theta_0)$, we can conclude that $\bm\theta_0$ also maximizes the negative classical cross entropy in the asymptotic regime:
\begin{equation}\label{eq:app_theta_0_H}
  \bm\theta_0 = \argmax_{\bm\theta \in \Theta} H \left( \Tr \left( \Pi_{Z} \rho \right) \| \Tr \left( \Pi_{Z} \sigma(\bm\theta) \right) \right).
\end{equation}
Hence the first step has been done. 
Now, as the second step, we use the standard consistency property of the maximum 
likelihood estimator $\hat{\bm\theta}_C$; that is, $\hat{\bm\theta}_C$ converges 
to $\bm\theta_0$ satisfying Eq.~(\ref{eq:app_theta_0_H}) and accordingly 
Eq.~(\ref{eq:app_theta_0_S}). 
Therefore, $\hat{\bm\theta}_C$ converges to $\bm\theta_0$ that is a solution to 
$\Tr \left( \rho \partial_{\bm\theta} \log \sigma(\bm\theta) \right) = \bm 0$.

We note that the above discussion assumes the measurement to be tomographically 
complete. 
However, even without the tomographically complete measurement, the convergence 
of the maximum likelihood estimator $\hat{\bm\theta}_C$ to the same $\bm\theta_0$ 
still holds if the measurement is chosen such that the condition 
(\ref{eq:app_eq_prob}) uniquely determines the state.

%%%%%%%%%%%%%%%%%%%%%%%%%%%%%%%%%%%%%%%%%%%%%%%%%%
\section{Brief review of quantum Fisher information}
\label{app:quntumFisherInfo}

Here we give a brief review of quantum Fisher information, including the symmetric 
logarithmic derivative Fisher (SLD) information and the Bogoljubov-Kubo-Mori (BKM) 
Fisher information. 
We especially derive a representation (\ref{eq:app_KMBFisher_alt}) of the BKM Fisher 
information, which is used in Eqs. (\ref{eq:LL_J_Q_est}) and (\ref{eq:shadow_hat_J_Q}).
For further details, we refer to 
Refs.~\cite{hasegawa1997Exponential,amari2000methods,hayashi2002Two}.

The classical Fisher information matrix (the $(i,j)$-element) is defined as
\begin{equation}
  g_{C;ij}(\bm\theta) = \mathbb{E}_{f(x|\bm\theta)} \left[ l_i(x|\bm\theta) l_j(x|\bm\theta) \right], \quad l_i(x|\bm\theta) f(x|\bm\theta) = \frac{\partial f(x|\bm\theta)}{\partial \theta_i},
\end{equation}
for a parametric model $\{ f(x|\bm\theta) : \bm\theta \in \Theta \subset \mathbb{R}^p \}$, 
where $l_i(\bm\theta)$ has another form
\begin{equation}
  l_i(\bm\theta) = \frac{\partial \log f(x|\bm\theta)}{\partial \theta_i}.
\end{equation}
We note that $g$ in this appendix represents a metric, although $g$ in the main article 
is used to represent the probability distribution.

There are several variants of quantum Fisher information, meaning that we thus need to 
choose a proper one depending on the purpose. 
One is the Symmetric Logarithmic Derivative (SLD) Fisher information defined by
\begin{eqnarray}
  g^{(s)}_{Q;ij}(\bm\theta) = \frac{1}{2} \Tr \left( \sigma(\bm\theta) ( L^{(s)}_i(\bm\theta) L^{(s)}_j(\bm\theta) + L^{(s)}_j(\bm\theta) L^{(s)}_i(\bm\theta) ) \right), \nonumber\\
  \frac{1}{2} \left( \sigma(\bm\theta) L^{(s)}_i(\bm\theta) + L^{(s)}_i(\bm\theta) \sigma(\bm\theta) \right) = \frac{\partial \sigma(\bm\theta)}{\partial \theta_i},
\end{eqnarray}
for a parametric quantum state $\{ \sigma(\bm\theta) : \bm\theta \in \Theta \subset \mathbb{R}^p \}$.
The SLD Fisher information appears in the quantum Cram\'er-Rao lower bound.

Another quantum analogue is the BKM Fisher information 
defined by
\begin{equation}\label{eq:app_KMBFisher}
  \fl g^{(b)}_{Q;ij}(\bm\theta) = \int_{0}^{1} \Tr \left( \sigma(\bm\theta)^t L^{(b)}_i(\bm\theta) \sigma(\bm\theta)^{1-t} L^{(b)}_j(\bm\theta) \right) dt, \quad \int_{0}^{1} \sigma(\bm\theta)^t L^{(b)}_i(\bm\theta) \sigma(\bm\theta)^{1-t} dt = \frac{\partial \sigma(\bm\theta)}{\partial \theta_i}.
\end{equation}
Importantly, $L^{(b)}_i(\bm\theta)$ can be represented as 
\begin{equation}\label{eq:app_logarithmicDerivative}
  L^{(b)}_i(\bm\theta) = \frac{\partial \log \sigma(\bm\theta)}{\partial \theta_i}.
\end{equation}
For example, see Section 7.3 of \cite{amari2000methods} for the derivation of this logarithmic derivative.
Also, $g^{(b)}_{Q;ij}(\bm\theta)$ has the following representation:
\begin{equation}\label{eq:app_KMBFisher_alt}
  g^{(b)}_{Q;ij}(\bm\theta) = \Tr \left( \sigma(\bm\theta) \partial_i \partial_j \log \sigma(\bm\theta) \right).
\end{equation}
To see this, let us consider the following equality: 
\begin{equation*}
  \Tr \left( \sigma(\bm\theta) \partial_i \partial_j \log \sigma(\bm\theta) \right) = \partial_i \{ \Tr \left( \sigma(\bm\theta) \partial_j \log \sigma(\bm\theta) \right) \} - \Tr \left( \partial_i \sigma(\bm\theta) \partial_j \log \sigma(\bm\theta) \right),
\end{equation*}
Note that $\Tr \left( \sigma(\bm\theta) \partial_j \log \sigma(\bm\theta) \right)$ in 
the first term of RHS is actually zero:
\begin{eqnarray*}
  \Tr \left( \sigma(\bm\theta) \partial_j \log \sigma(\bm\theta) \right) &= \Tr \left( \sigma(\bm\theta) L^{(b)}_j(\bm\theta) \right) \\
  &= \int_{0}^{1} \Tr \left( \sigma(\bm\theta)^t L^{(b)}_j(\bm\theta) \sigma(\bm\theta)^{1-t} \right) dt \\
  &= \Tr(\partial_j \sigma(\bm\theta)) = 0.
\end{eqnarray*}
The first equality uses Eq.~(\ref{eq:app_logarithmicDerivative}) and the third equality is obtained by Eq.~(\ref{eq:app_KMBFisher}).
This leads to
\begin{eqnarray*}
  \Tr \left( \sigma(\bm\theta) \partial_i \partial_j \log \sigma(\bm\theta) \right) &= - \Tr \left( \partial_i \sigma(\bm\theta) \partial_j \log \sigma(\bm\theta) \right) \\
  &= g^{(b)}_{Q;ij}(\bm\theta).
\end{eqnarray*}
The last equality uses Eq.~(\ref{eq:app_KMBFisher}), and thus we arrive at 
Eq.~(\ref{eq:app_KMBFisher_alt}).

An interesting feature is that the BKM metric appears in the limit of the quantum 
relative entropy \cite{hayashi2002Two}. 
This justifies the use of the BKM metric in the proposed information criteria. 
Moreover, in terms of information geometry, only the BKM metric introduces a dually 
flat structure for a family of density matrices, which is the novel property that 
the classical Fisher information has for a family of probability distributions. 
As an application, the BKM metric is chosen to develop the quantum generalization 
of the classical mirror descent algorithm \cite{sbahi2022Provably}.

%%%%%%%%%%%%%%%%%%%%%%%%%%%%%%%%%%%%%%%%%%%%%%%%%%
\section*{References}
\bibliographystyle{iopart-num}
\bibliography{references}

\end{document}